\documentclass[preprint]{aastex}
\usepackage{epsfig}
\received{}

\revised{}
\accepted{}
\ccc{}
\cpright{}{}

\shorttitle{Dust Formation in the young core-collapse supernova remnant E0102 }
\shortauthors{}

\newcommand{\oivf}{\ion{[O}{4}]}
\newcommand{\siif}{\ion{[S}{2}]}
\newcommand{\siiif}{\ion{[S}{3}]}
\newcommand{\siliif}{\ion{[Si}{2}]}

\newcommand{\neiif}{\ion{[Ne}{2}]}
\newcommand{\neiiif}{\ion{[Ne}{3}]}

\newcommand{\nevf}{\ion{[Ne}{5}]}

\newcommand{\ariif}{\ion{[Ar}{2}]}

\newcommand{\feiif}{\ion{[Fe}{2}]}

\newcommand{\mic}{$\mu$m}

\newcommand{\spitzer}{\textit{Spitzer}}

\def\e0102{1E 0102.2-7219}

\begin{document}

\title{{\it Spitzer} Observations of the Young Core-collapse Supernova Remnant
1E0102-72.3: Infrared Ejecta Emission and Dust Formation}

\author{J. Rho\altaffilmark{1}, W. T. Reach\altaffilmark{1}, A.
Tappe\altaffilmark{1, 2}, U. Hwang\altaffilmark{3}, J. D.
Slavin\altaffilmark{2}, T. Kozasa\altaffilmark{4}, L. Dunne\altaffilmark{5} }
\affil{$^1$ Infrared Processing and Analysis Center, California Institute of
Technology, Pasadena, CA 91125; rho@ipac.caltech.edu}
\affil{$^2$ Harvard-Smithsonian Center for Astrophysics, MS 83, 60 Garden
Street, Cambridge, MA 02138}
\affil{$^3$ NASA Goddard Space Flight Center, Greenbelt, MD 20771}
\affil{$^4$ Department of Cosmosciences, Graduate School of Science,  Hokkaido
University, Sapporo 060-0810, Japan}
\affil{$^5$ School of Physics and Astronomy, University of Nottingham,
University Park, Nottingham, NG7 2RD, UK
}
\begin{abstract}
We present \spitzer\ IRS and IRAC observations of the young supernova
remnant E0102 (SNR 1E0102-7219) in the Small Magellanic Cloud.  The
infrared spectra show strong lines of Ne and O, with the \neiif\ line at
12.8 \mic\ having a large velocity dispersion of 2,000-4,500 km s$^{-1}$
indicative of fast-moving ejecta. Unlike the
young Galactic SNR Cas A, E0102 lacks emission from Ar and Fe. 
Diagnostics of the observed [Ne III] 
line pairs imply that [Ne III] emitting ejecta have a low temperature of 650
K, while [Ne V] line pairs 
imply that the infrared [Ne V] emitting ejecta have a high density of
$\sim$10$^{4}$ cm$^{-3}$.  We have calculated radiative shock models for
various velocity ranges including the effects of photoionization. The
shock model indicates that the \nevf\ lines come mainly from the cooling
zone, which is hot and dense, whereas [Ne~II] and [Ne~III] come mainly
from the photoinization zone, which has a low temperature of 400 - 1000
K. We estimate an infrared emitting Ne ejecta mass of 0.04 $M_\odot$ from the infrared
observations, and discuss implications for the progenitor mass.  The
spectra also have a dust continuum feature peaking at 18 \mic\ that
coincides spatially with the ejecta, providing evidence that dust
formed in the expanding ejecta.
The 18 \mic-peak dust feature is
fitted by a mixture of MgSiO$_3$ and Si dust grains, while the rest of
the continuum requires either carbon or Al$_2$O$_3$ grains.   We measure the total dust mass
formed within the ejecta of E0102 to be $\sim$0.014
M$_\odot$.
The dust mass in E0102 is thus a factor of a few smaller than that in
Cas A.  The composition of the dust is also different, showing
relatively less silicate and likely no Fe-bearing dust, as is suggested
by the absence of Fe-emitting ejecta.

\end{abstract}
\keywords{supernovae:general - infrared: general  - dust:ISM - supernova remnants:SNR 1E0102-72.3}
\clearpage

\clearpage
\section{Introduction}

Young supernova remnants (YSNRs) are the most viable astrophysical
laboratories for the study of dust formation, nucleosynthesis of heavy
elements, cosmic-ray acceleration, and shock physics.  Of these, dust
formation is the least well-studied despite its invocation in efforts
to explain some of the most important problems in astrophysics,
including the isotopic anomalies of heavy elements in meteorites
(Clayton 1982), dust formation in the early Universe \citep{noz03}, 
the measured abundances of Galactic cosmic rays (Ellison, Drury
\& Meyer 1997), and the interstellar dust budget crisis \citep{jon94}.

Meteoritic and astronomical studies show that presolar, cosmic grains
condense in  the dense, warm stellar winds of evolved stars  and
in the ejecta of supernovae.     Mantles of the pre-existing dust in
molecular clouds  are vaporized as the forming stars and planetary
systems heat them. A small fraction of the dust survives solar system
formation without alteration, protected inside asteroids.  The most
abundant presolar grains are SiC, nanodiamonds, amorphous silicates,
fosterite and enstatite,  and corundum (Al$_2$O$_3$) \citep{messenger}.
Some isotopic anomalies of heavy elements in meteorites have been
attributed to the dust that had condensed deep within expanding
supernovae and some have the r-process composition commonly
associated with Type II SNe \citep[][references therein]{cla97,
  cla04}.

Recent deep sub-mm observations have also shown there to be galaxies
and QSOs with very large dust masses ($>10^8 ~\rm{M_{\odot}}$) at $z>5$
(Wang et al. 2008; Beelen et al. 2006; Bertoldi et al. 2003). The
timescales for low mass (AGB) stars to release their dust are too long to explain
these high redshift systems (Morgan \& Edmunds 2003). In contrast,
supernovae evolved from massive stars produce copious amounts of heavy elements and release them
on short timescales. Theoretical modeling of the conditions in the
supernova ejecta indicates that Type-II SNe are sources of dust
formation and should produce substantial quantities of dust,
on the order of a solar mass per explosion \citep{den03,den06,tod01,noz03}.  Yet until very recently, there
existed little observational evidence that this actually occurs.

There is now sufficient evidence for dust formation in core-collapse
supernovae.  For SN 1987A, this includes dust emission, dust absorption
and a drop in line intensities for the refractory elements that
signals that dust is being formed \citep{lucy89,lucy91}. 
Formation of dust in the expanding ejecta of SN 1987A was explored theoretically 
\citep{kozasa89,kozasa91}.
The Type II-P SN1999em also
showed clear signs of dust formation (Elmhamdi et al. 2003).
Detections of CO fundamentals with \spitzer\ observations of SN 2005af
and Cas A further provide an indirect evidence of dust formation in
SNe \citep{kot06,rho09}.

For the young Galactic SNR, Cas\,A, ISO observations gave evidence for
the association of the dust with the ejecta by requiring a mixture of
dust grains that are not typical of the ISM, while showing strong
spatial correlations between the high-velocity infrared emission
lines, the dust continuum emission, and the optically emitting ejecta
(Lagage et al.\ 1996 and Arendt et al.\ 1999). Submillimeter
observations of Cas\,A and Kepler with SCUBA \citep{dun03, mor03a}
suggest the presence of large amounts of cold dust ($\sim
0.3-2\,\rm{M_{\odot}}$ at 15--20 K), but with some controversy related
to the presence of foreground material.  \cite{kra04} showed that much
of the 160 $\mu$m emission observed with Multiband Imaging Photometer
for {\it Spitzer} (MIPS) is foreground material, suggesting there is
no cold dust in Cas A. \cite{wil05}, however, used CO emission towards
the remnant to show that up to about a solar mass of dust could still
be associated with the ejecta rather than with foreground material.
New submm polarimetry data confirms that a significant fraction 
($>$30\%) of the submm flux originates from within the remnant 
(Dunne et al. 2009).
The case for dust formation in supernova ejecta was
strengthened by {\it Spitzer} observations of Cas A, for which
line-free dust maps were found to remarkably resemble the infrared
ejecta-line maps of [Ar~II], [O~VI], and [Ne~II]; the estimated dust
mass is between 0.02 and 0.054 M$_{\odot}$ \citep{rho08}.
A similar dust mass is reproduced by 
theoretical models of dust formation that include dust destruction and stochastic heating  
\citep{kozasa09} and
dust destruction \citep{bianchi07,bianchi09} for Cas A.

\begin{figure}
\plotone{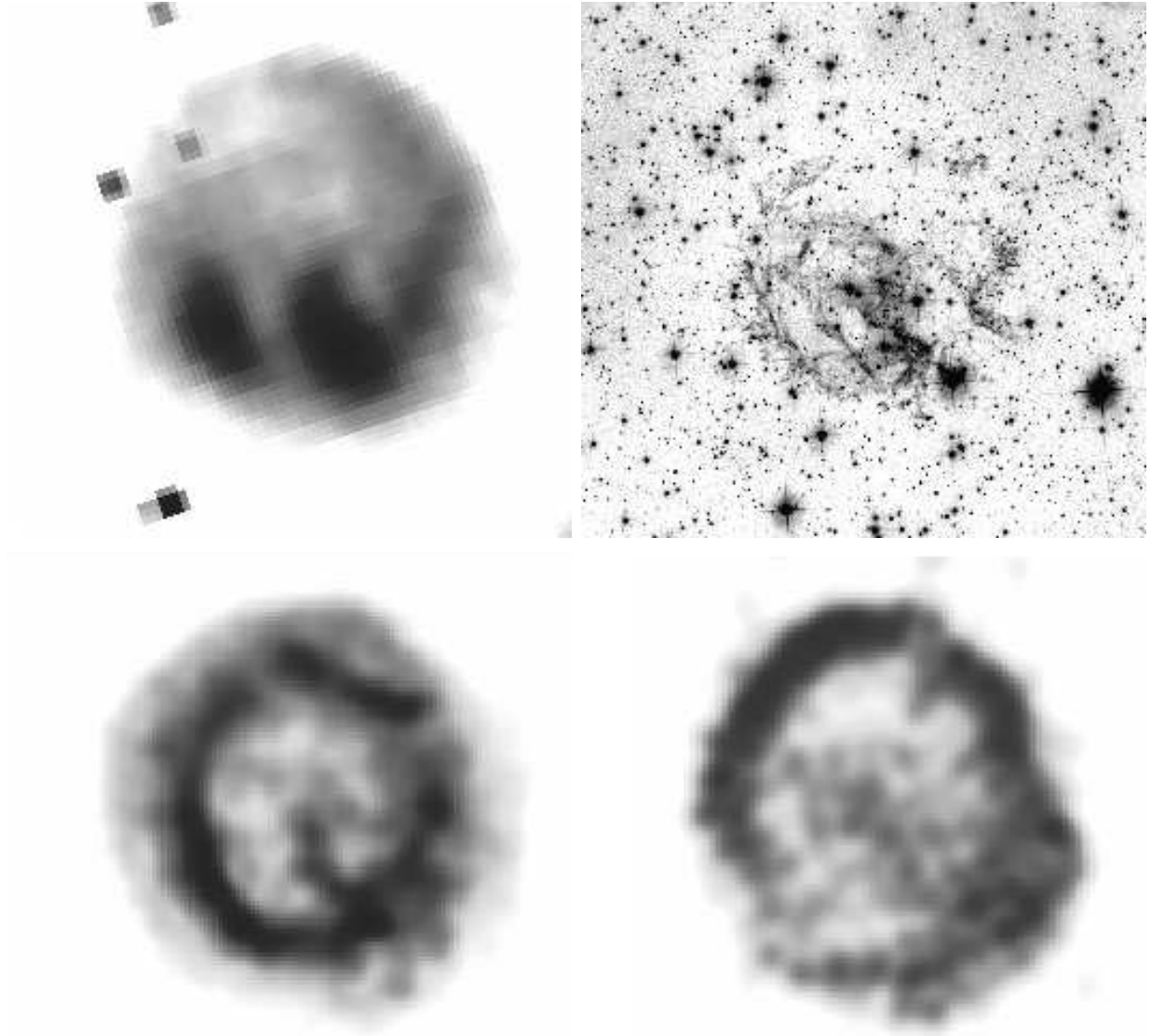}
\caption{ Four images of E0102 on the same angular scale (clockwise from upper left):
    infrared MIPS
  24 \mic\ \citep{sta06}, HST \citep{fin06}, 
ACTA radio at 6 cm \citep{amy93}, and
{\it Chandra} X-rays
  \citep{fla04}. The brightest
  ring of radio emission traces the forward shocked material, and the brightest
  ring of X-ray emission traces the reverse-shocked ejecta. }
\label{multi}
\end{figure}

Another excellent source to study dust formation in SNe is the young
supernova remnant (SNR) 1E 0102-7219 (SNR B0102-72.3; E0102
hereafter) in the Small Magellanic Cloud.  As summarized in Table
\ref{tproperties}, it is located at R.A.\ $01^{\rm h} 04^{\rm m}
2.57^{\rm s}$ and Dec.\ $-72^\circ$01$^{\prime} 52.3^{\prime \prime}$
(J2000), and has a diameter of 44", which corresponds to $\sim$ 13.04
pc for a distance of 61.1 kpc to the SMC \citep{hil05,kel06}.
Various wavelength images of E0102 are shown in Figure \ref{multi}.
Highly enriched abundances of oxygen and neon relative to carbon and
magnesium (Blair et al. 2000), as is characteristic of massive
progenitors, place E0102 in the class of $``$oxygen-rich" SNRs, along
with Cas~A.  The optical ejecta in E0102 have substantial velocities
that exceed 1000 km s$^{-1}$ and imply a dynamic age of about 1000 yr
\citep{tuo83}.  The X-ray proper motion of the blast wave measured by
Hughes et al.  (2000) is also consistent with this age. More recent
HST measurements of the ejecta filament proper motions imply a
kinematic age of 2050$\pm$600 yr that is a factor of two higher
(Finkelstein et al.  2006), but the proper motion measurements are
consistent within their quoted errors.  The composition of the
optically emitting ejecta suggests a massive progenitor that underwent
either a Type Ib \citep{bla00} or Type Ic \citep{fla04} supernova
explosion.  The optical expansion rates place E0102 currently in
transition from free expansion to the Sedov phase.  The low extinction
toward E0102 compared with that of Galactic SNRs makes it possible to
observe the infrared (IR) emission without contamination by emission from clouds
along the line of sight (as seen, for example, in Cas A,
\cite{kra04}).  The low metallicity environment of the SMC also makes
E0102 a useful Type II 
prototype to compare with SNRs at higher redshifts.

In the infrared, the 24 $\mu$m \spitzer\ MIPS image of E0102 presented
by Stanimirovic et al. (2006) shows a filled morphology with two
prominent elongated filaments that resemble structures seen in the
X-ray image; the remnant was not detected at 8 or 70 \mic.  The
infrared emission appears to be mainly associated with 
reverse shocks of the hot X-ray gas.   
An IR emission peak around
24 $\mu$m would suggest the presence of hot dust in E0102 with
T$\sim$120 K, with most of the dust being centrally located, based on the
24 $\mu$m image.  Assuming the dust originates in the ISM and using
typical ISM dust absorption coefficients \citep{dra84}, the estimated
dust mass is $8\times 10^{-4} M_\odot$.  Even if all the dust were actually
formed in the explosion, this dust mass is lower than might be expected based
on some theoretical models. Most importantly, since the 24 \mic\ emission
contains both line and continuum, 
not photometry but spectroscopy is required to estimate the mass and composition
of dust as was demonstrated by \cite{rho08}. 

In this paper, we present \spitzer\ spectral and imaging observations
of E0102 with IRS and IRAC. We find that the infrared ejecta emission
is dominated by Ne and O and
detect a dust feature peaking at 18 \mic\ which is spatially
coincident with the ejecta emission.  
A high portion of the dust clearly comes from ejecta, rather
than being all associated with circumstellar/interstellar material.
We show that some of dust is at lower temperature and a total dust mass is higher 
than inferred from earlier observations.
We discuss physical conditions of infrared emitting Ne and its inferred ejecta mass using 
emission line diagnostics and shock models.

\section{Observations}

\begin{figure}
\epsfig{figure=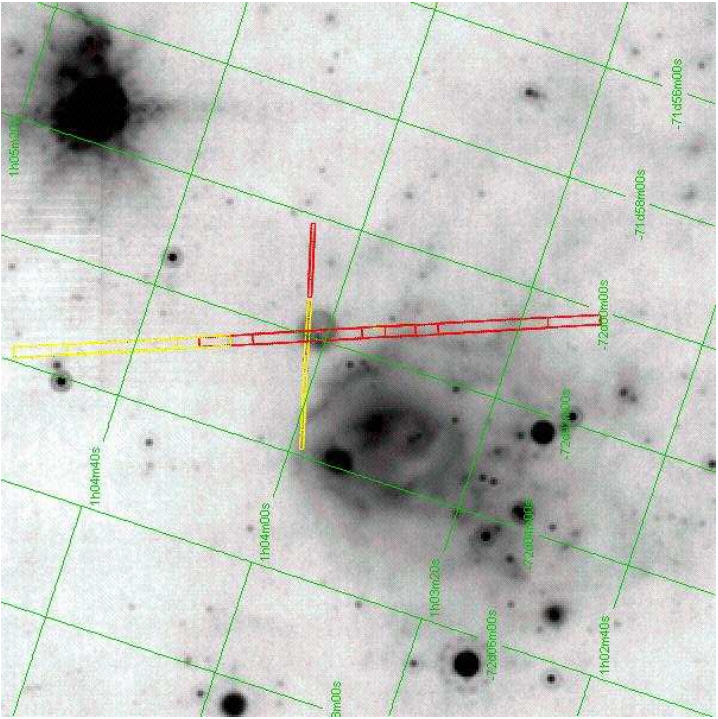,angle=90,height=9.5truecm}
\caption{IRS slit coverage is superposed on a MIPS 24$\mu$m image \citep{sta06} 
of E0102. North is to the right and east is up. 
The large diffuse source located southwest of E0102 is
the [H~II] regions of N76.}
\label{e0102irsslit}
\end{figure}

We performed an IRS staring observation toward the southeastern shell
of E0102 (R.A.\ $01^{\rm h} 04^{\rm m} 04.04^{\rm s}$ and
Dec.\ $-72^\circ$02$^{\prime} 00.5^{\prime \prime}$, J2000, see Figure
\ref{e0102irsslit}) as a part of our Young SNR \spitzer\ GO program
(PI: Rho).  The Long Low (LL: 15-40 $\mu$m) IRS data were taken on
2005 August 14 with 6 cycles of 30 sec exposure time; this yields a
total exposure time of 360 sec for the first and second staring
positions.  The Short Low (SL: 5-15 $\mu$m) IRS observations were made
with 3 cycles of 60 sec exposure time and one cycle covers 2 dither positions; this yields a total exposure
time of 360 sec per sky position. 

\begin{figure}[!h]
\psfig{figure=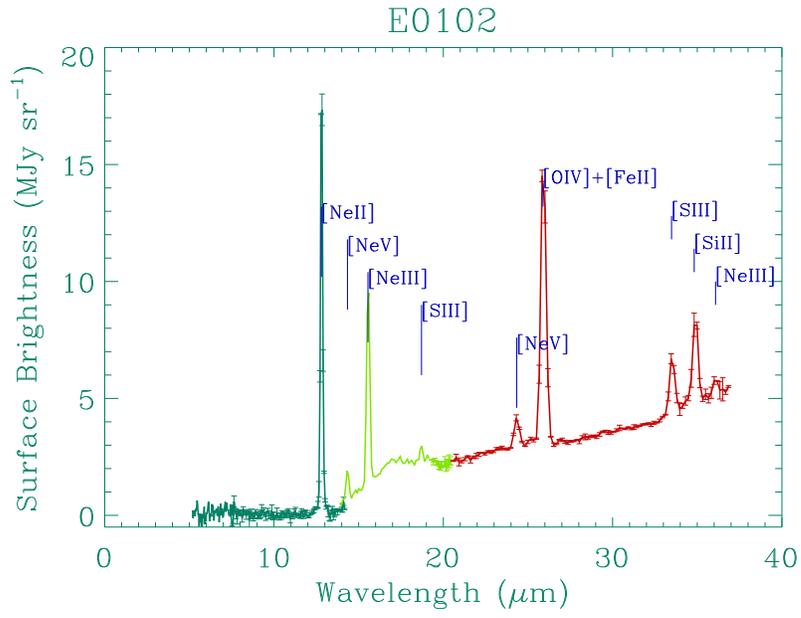,angle=0,height=8.5truecm}
\caption{{\it Spitzer} IRS spectrum of E0102.}
\label{E0102irs}
\end{figure}

\begin{figure}
\epsfig{figure=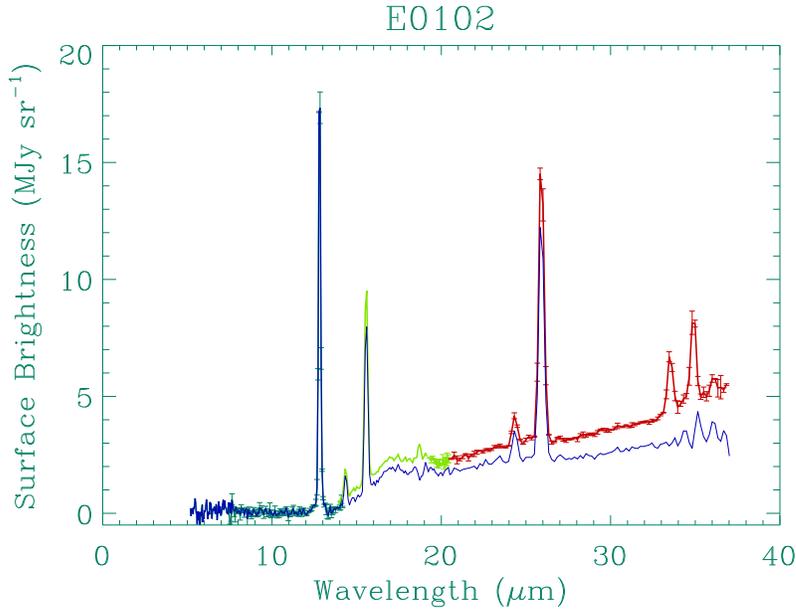,angle=0,height=8.5truecm}
\caption{
The spectrum before (top),  and the spectrum (bottom) after,
local-background subtraction.}
\label{E0102irsb}
\end{figure}

\begin{figure}
\vbox{
\epsfig{figure=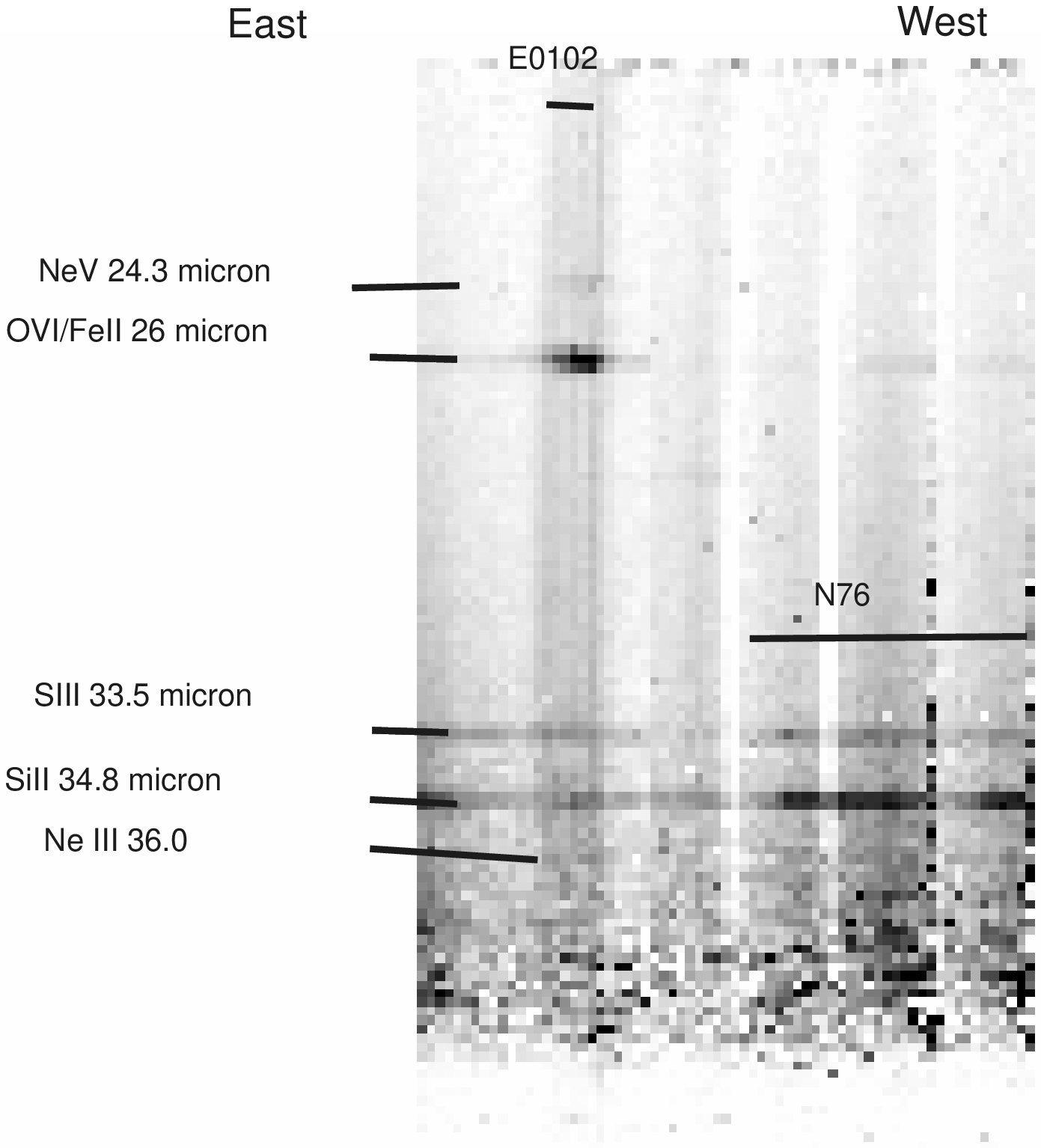,angle=0,height=8truecm}
\epsfig{figure=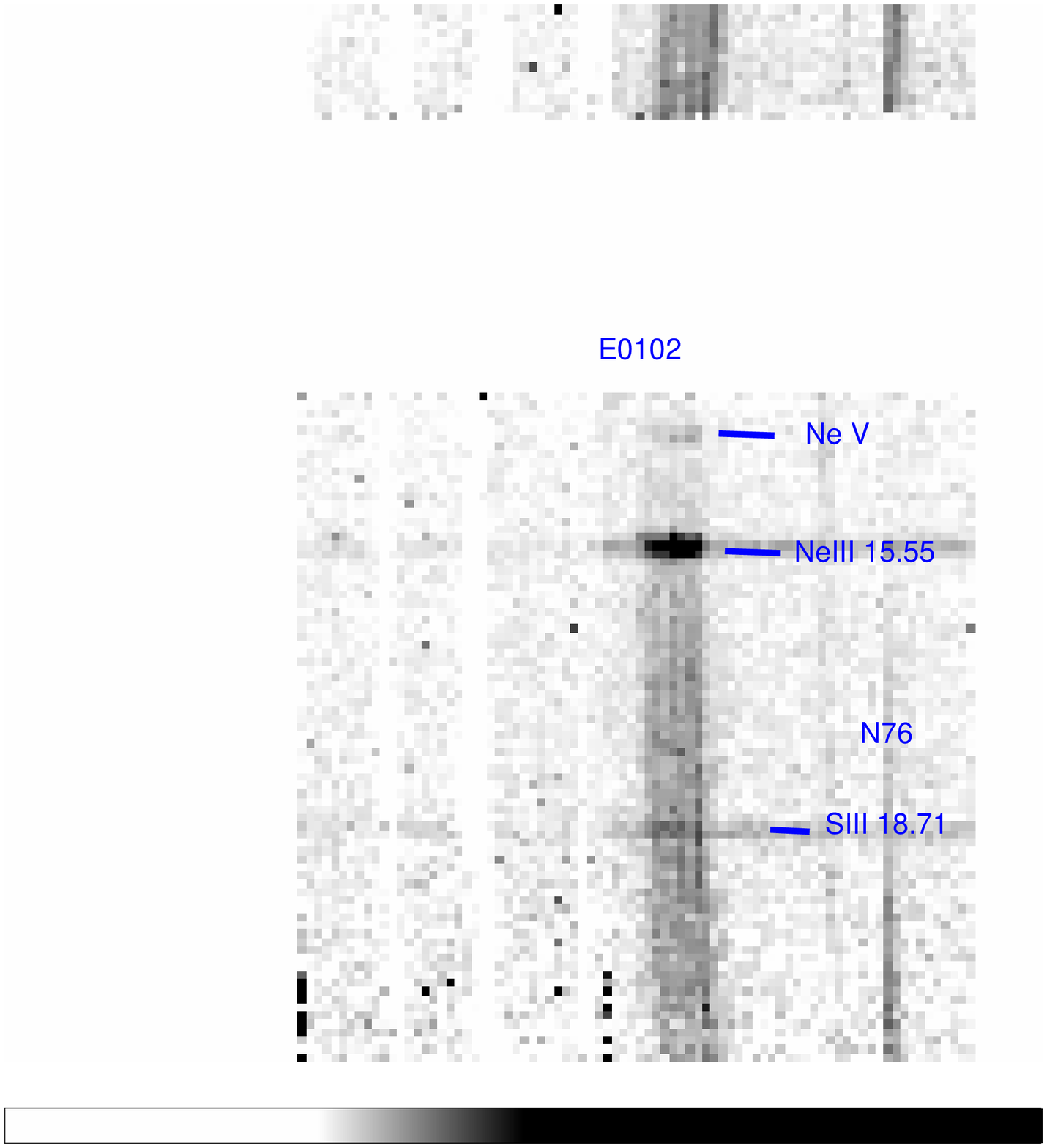,angle=0,height=8truecm}
}
\caption{Merged LL position-wavelength images (left: LL2, right: LL1); X-axis is
spatial direction along the slit covering E0102 and N76 and y-axis is wavelength.
The image shows that E0102 emits continuum, [Ne~V], [Ne~II], and [O~IV] lines.}
\label{pwLL}
\end{figure}

The IRS spectra were processed using the S15.3 pipeline products as
follows. First, we removed rogue pixels (bad pixels mainly due to
cosmic rays) by subtracting nearby empty sky fields with the same
exposure time. We took the empty sky data from our observation in the
alternate orders or from archival data (AORs 14706944, 14707712,
14708480, and 14708824).  Any remaining rogue pixels were removed by
using the SSC tool
IRSCLEAN\footnote{http://ssc.spitzer.caltech.edu/archanaly/contributed/}.
Second, we extracted the spectrum using SPICE, and removed fringes.
Third, we applied aperture and
slit-loss corrections for diffuse emission (also see Tappe, Rho, \&
Reach 2006).  The reduced spectrum is shown in Figure
\ref{E0102irs}. Additionally we subtracted a local background spectrum
using the relatively emission-free regions right outside the SNR in
our slit data (specifically, the region outside the eastern side of
E0102 at RA. 16.060, Dec.  -72.039).  The background-subtracted
spectrum is shown in Figure \ref{E0102irsb}. 
Comparison of the two spectra is described in \S3.1. 
 The staring mode yields 4 coadded spectral
images (positions 1 and 2, LL1 and LL2) which we merged to make a
position--wavelength map as shown in Figure \ref{pwLL}.  The location
of E0102 on this map is marked by the appearance of both continuum
emission and strong \neiiif\ (15.55 \mic) and weak \nevf\ (23.3 \mic)
emission lines.

We also used archival IRS mapping data from the IRS Legacy SMC program
(PI: Bolatto). These maps were made for large regions of the SMC and
included E0102. The observations used 1 cycle of 30 sec exposure time
with 5 arcsec overlap, yielding a total of 60 sec per sky
position. These mapping data were 6 times shallower than our GO IRS
staring mode data.

We made a deep IRAC observation as a part of our GO program.  The IRAC
observations were made for a full array with 30 sec frame time and a 9
position random dither pattern, yielding 36 sec exposure per sky
position.  The data were taken on 2005 June 12. IRAC bands cover the
wavelength ranges 3.2--4.0, 4.0--5.0, 5.0--6.4, and 6.4--9.4 \mic,
respectively, with a FWHM of 1.7\arcsec.

\section{Results}

\subsection{IRS Spectra: Continuum and Ejecta Lines}

The IRS spectra of E0102 from the southeastern shell in Figure
\ref{E0102irs} show both line and continuum emission.  The detected
lines include [Ne~II] (12.8 $\mu$m), [Ne~V] (14.3 $\mu$m), [Ne~III]
(15.5 $\mu$m), [Ne~V] (24.3 $\mu$m), [O~IV] and/or [Fe~II] (26 $\mu$m),
and [Ne~III] (36.07 $\mu$m).  The local-background subtraction
effectively removed the [S~III] (33.5 $\mu$m), [Si~II] (34.8 $\mu$m),
and [S~III] (18.7 $\mu$m) lines, but this background is likely to be
contaminated by the nearby \ion{H}{2} region of N76.
When we generated
a [Si~II] map from the IRS cube,  E0102 appeared 
as a  [Si~II]  emitter. 
The lack of clean local background
data thus makes it unclear whether there are actually any S and Si
lines associated with E0102.  The position-wavelength LL maps in
Figure ~\ref{pwLL} show that the SNR E0102 emits continuum emission,
\nevf, \neiiif, and \oivf (or [Fe~II]).  The 26 $\mu$m blend is likely
to be dominated by [O~IV], and unlikely to contain any [Fe II],
because other [Fe ~II] lines such as those at 17.9, 6.7, and 5.3 $\mu$m
are not detected.  The infrared spectrum of E0102 is thus dominated by
Ne and O lines.

The detected lines and their brightnesses are summarized in Table
~\ref{table:tlineflux}.  We measured the line brightnesses using two methods:
a Gaussian fitting, and integration of line fluxes  over wavelength
ranges. The latter is required in order to account for red- and
blue-shifted velocity dispersions. To obtain unreddened line
brightness, we used an extinction value (Av) of 0.08 mag \citep{bla89},
which is equivalent to 2 ($<$8)$\times$ 10$^{20}$ cm$^{-2}$. This value
is consistent with those of X-ray measurements \citep{sas06}. The
extinction corrected values of A$_\lambda$ are also shown in Table
~\ref{tproperties}. Because the extinction is very small, the unreddened
line brightnesses differ by less than 0.3\%.

The E0102 spectrum shows a dust feature between 16 and 20 \mic\ peaking
at 18.5 \mic\ (we call it the 18 \mic-peak dust feature, hereafter).
This feature occurs in the wavelength range of a prominent silicate
feature, but it is very broad and extends to shorter wavelengths. 
We present a multi-composition fit in
\S4.1.

\subsection{IRAC images}
\begin{figure}
\psfig{figure=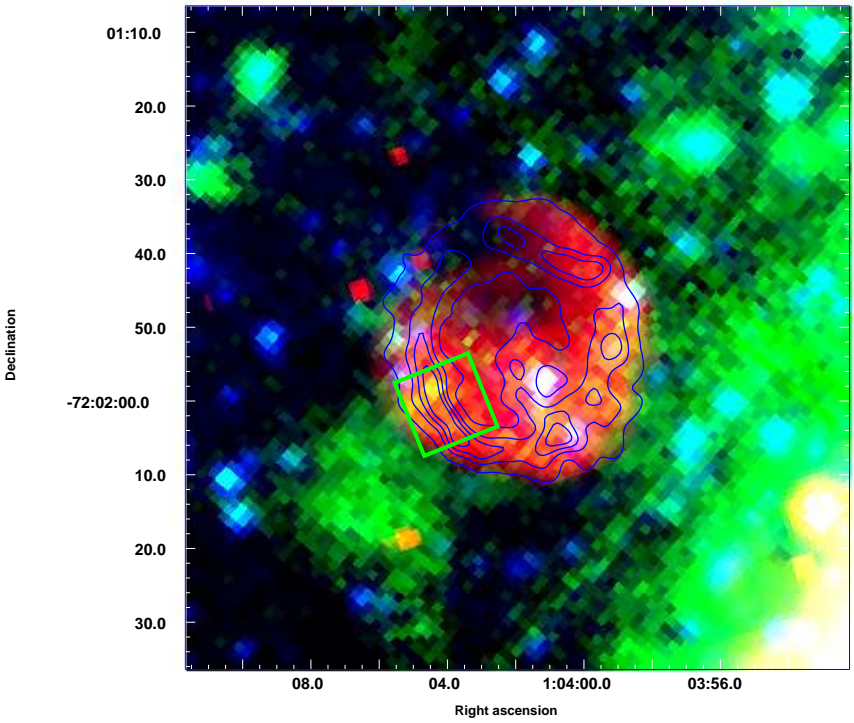,angle=0,height=10truecm}
\caption{Mosaicked three-color infrared images of E0102. Red: MIPS
24 \mic\ (from Stanimirovic et al. 2006); Green: IRAC 8 \mic; Blue: IRAC
5.8 \mic\ images.  The IRAC 8\mic\ flux ranges from 0.15 to 0.2 MJy
sr$^{-1}$. Possible detection of faint 8\mic\ emission (seen in
yellow) from the southeastern shell is suggested.  The IRS slit
coverage is marked as a green box.  Note that the remnant is not
detected at 5.8 \mic. 
The bright emission at the bottom right of the image is from N76. 
X-ray contours \citep{gae00} are superposed.
}
\label{E0102irac}
\end{figure}

Faint emission, particularly from the southeastern shell, is seen in
our IRAC 8 $\mu$m image of E0102; the 8 $\mu$m emission corresponds
well with the MIPS 24 $\mu$m emission, as shown in Figure
\ref{E0102irac}. The flux range in the image is 0.15-0.2 MJy
sr$^{-1}$.  However, there is also some 8 \mic\ polycyclic aromatic hydrocarbons  (PAH) emission from
the nearby \ion{H}{2} region of N76 to the southwestern side of the
SNR and a large patch of diffuse emission outside the southwestern
shell of the SNR.
We did not detect 8 $\mu$m emission with IRS, but IRS has poorer
sensitivity than IRAC, with a limit of 0.24 MJy sr$^{-1}$ for our IRS
staring observation compared to 0.15 MJy sr$^{-1}$ for IRAC.  Because
the IRAC flux is measured relative to the nearby background, we
believe that the IRAC 8 \mic\ emission detected towards E0102 comes
from the SNR.  Moreover, dust emission is expected in the ejecta and
any emission from E0102 at 8 \mic\ is likely to be continuum emission.
A deep IRAC image, or a more sensitive observation with future
missions may be helpful in unambiguously determining if E0102 does
emit at 8 \mic.  No emission of E0102 at 3.6, 4.5 and 5.8 $\mu$m is
detected.

\subsection{Line and Dust Maps}

\begin{figure}
\plotone{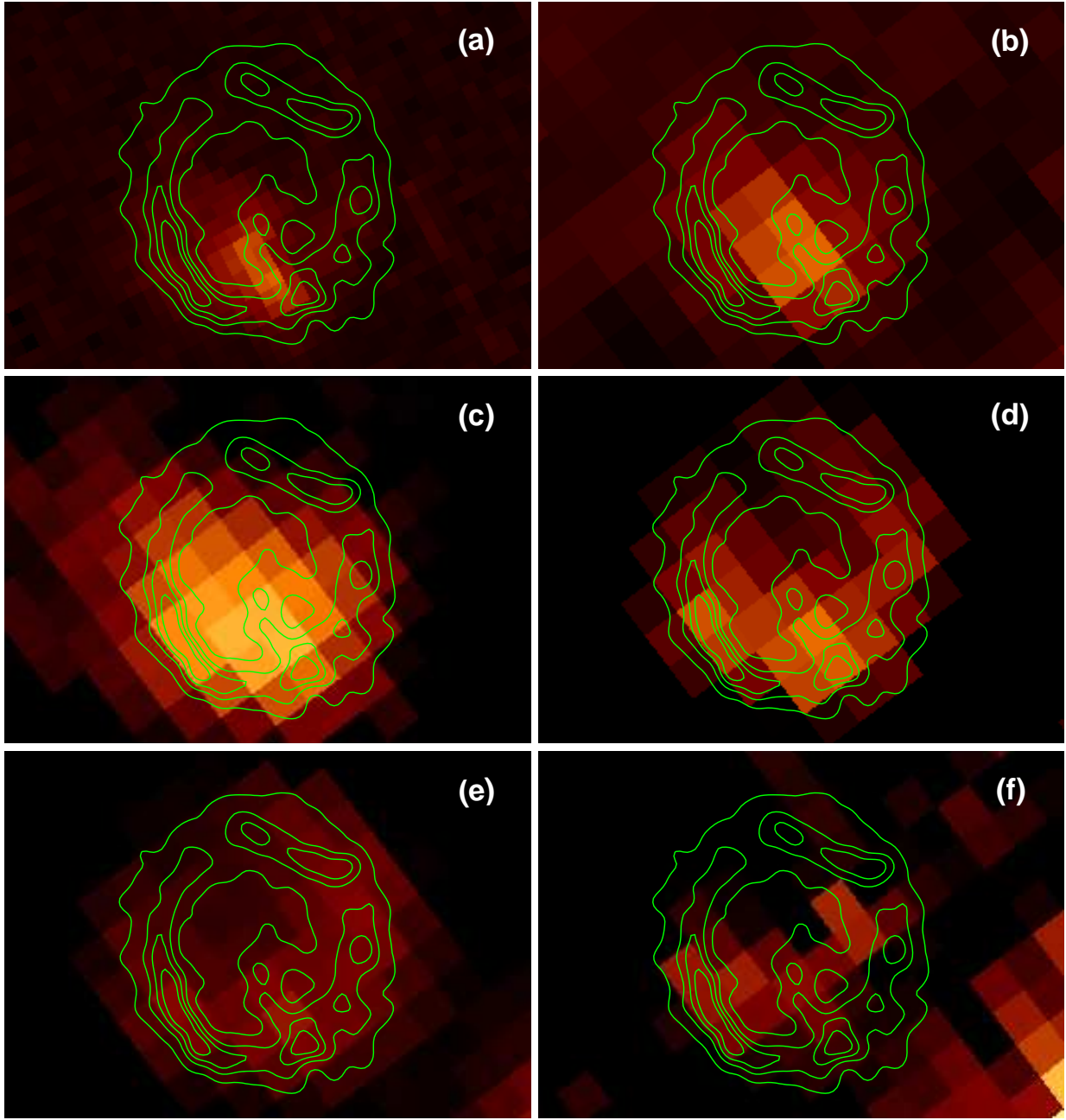}
\caption{ Line map images of E0102: (a) [Ne II] 12.8 $\mu$m, (b)
  [Ne~III] 15.55 $\mu$m, (c) \oivf+[Fe~II] 26 $\mu$m, (d) continuum
  16.5-20.5 \mic, (e) continuum 27-31 $\mu$m, and (f) [Si~II]
  34.8 $\mu$m.  X-ray contours \citep{gae00} are superposed.
The spatial resolution of [Ne II] image (from SL) is $\sim$2$''$   
and of the other images (from LL) is $\sim$5.1$''$.}
\label{E0102linemaps}
\end{figure}

\begin{figure}
\epsfig{figure=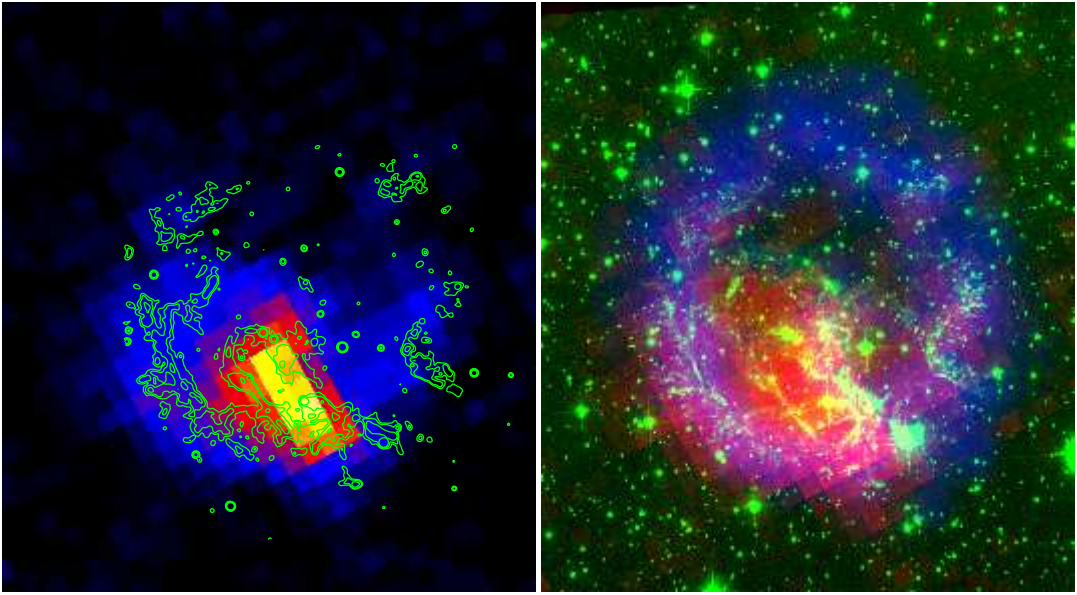,height=8truecm}
\caption{(left) \spitzer\ [Ne~II] intensity map (colors are from
yellow, red and blue in strength of brightness) overlaid with  HST [O~III] contours. (right) 
Color images of X-ray (blue), optical [O~III] (green) and infrared  [Ne II] 12.8$\mu$m (red) emission.}
\label{nexrayhstcontours}
\end{figure}

We generated line and continuum maps using the IRS mapping data, which
covered the SNR E0102 and the nearby \ion{H}{2} region N76.  The maps
are shown in Figure \ref{E0102linemaps}.  The line maps were generated
for \neiif, \neiiif, \oivf\ and \siiif\ by subtracting the continuum
baselines at neighboring wavelengths.  
 Infrared ejecta emission from \neiif, \neiiif, \oivf\ and
\siliif\ (Figure \ref{E0102linemaps}) coincides with the optical
         [O~III] ejecta emission as well as with X-ray ejecta
         emission.
In spite of the lower angular resolution for the infrared images, this
can be seen clearly in the overlay of the \neiif\ image with the
optical [O~III] contours (from Blair et al. 1989) and the X-ray {\it
  Chandra} image \citep{gae00} in Figure \ref{nexrayhstcontours} .
The infrared emission is strong at the southeastern shell and center,
both of which coincide with ejecta emission seen in X-ray and optical
images.  The infrared line emission is weaker in the northern and
northwestern regions, which are, however, also associated with X-ray
emitting ejecta.  The ejecta line profile of the northwestern region
shows significant line-broadening (see \S3.4 for details). 
 The
infrared continuum image 
is most similar to 
the infrared \oivf+\feiif\ (at 26 $\mu$m) line image, 
except in the northwestern shell.
However, this continuum emission at the northwestern
shell nicely follows the X-ray ejecta shell as shown in Figure
\ref{E0102linemaps}d, and is located inside the forward shocked
material traced by the radio shell (see Figure \ref{multi}).  The dust
emission is thus seen to be spatially correlated with the ejecta
emission.

\begin{figure}
\vbox{
\epsfig{figure=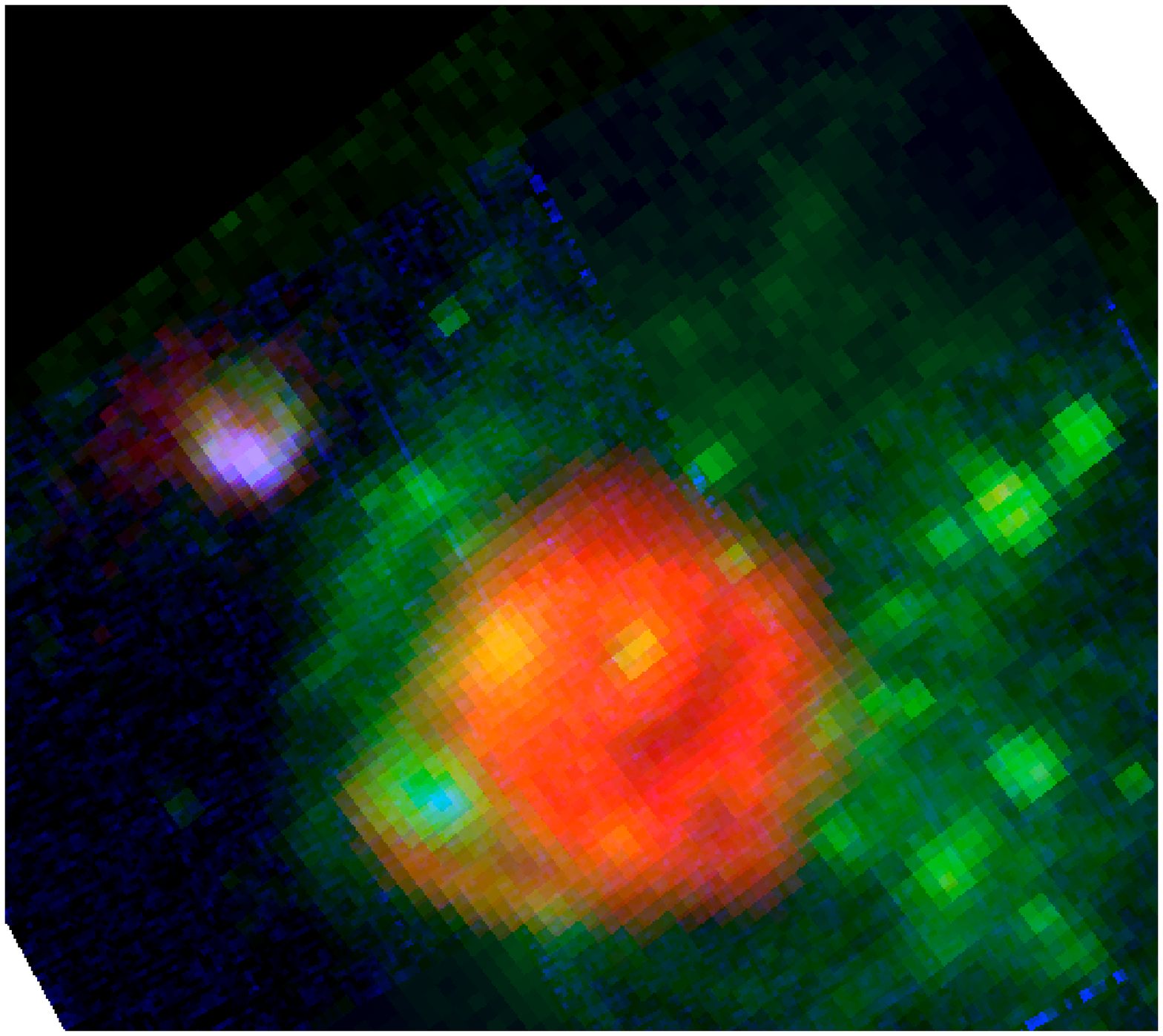,height=6.5truecm}
\epsfig{figure=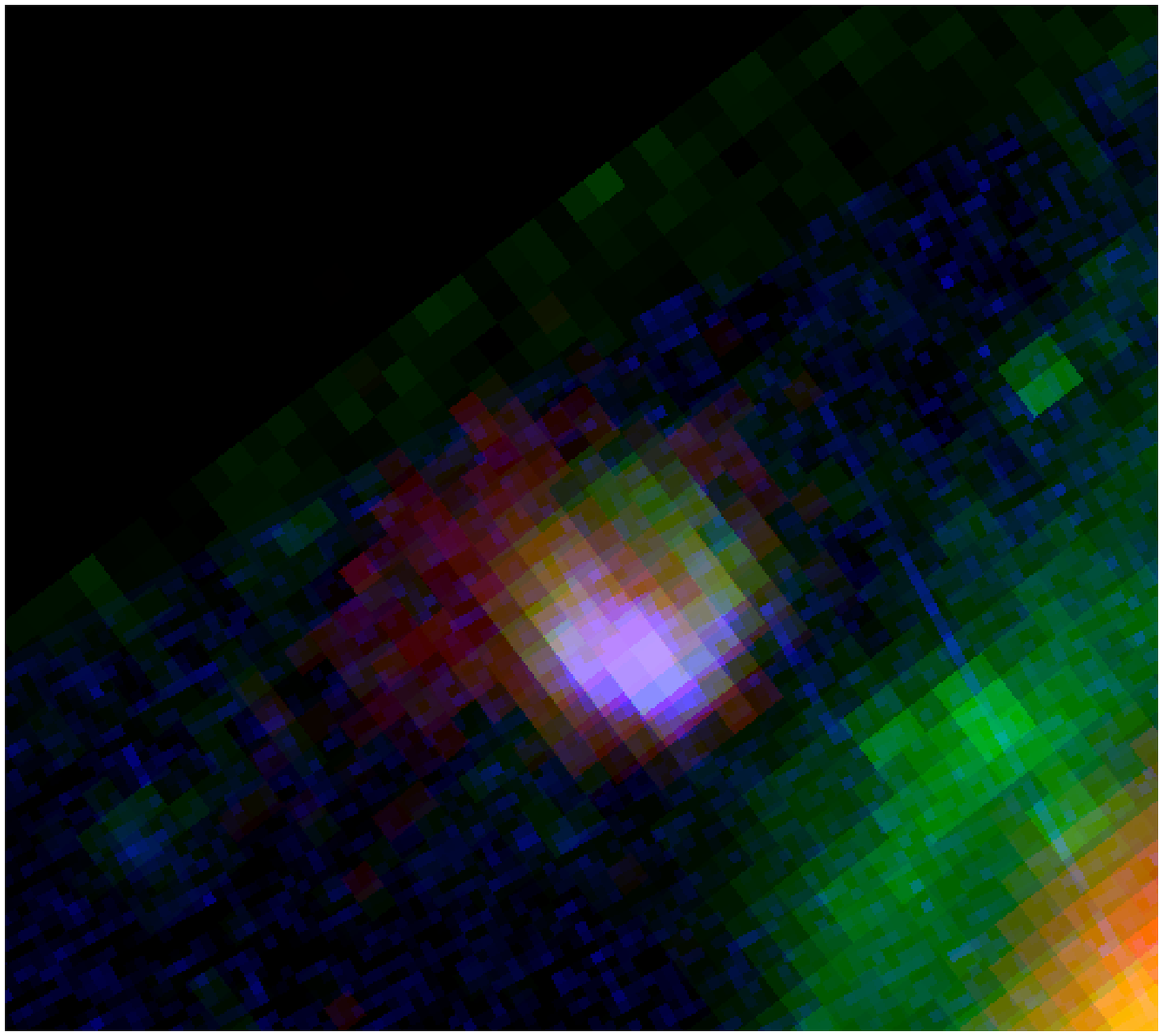,height=6.5truecm}
}
\caption{Mosaicked three-color images of (a) E0102 and N76, (b) E0102
  only: \neiif\ (blue), 16.5-20.5 \mic\ continuum (green) and \oivf\
(red). The image on the left is centered on R.A.= $01^{\rm h} 03^{\rm m}
45.6^{\rm s}$ and Dec.= $-72^\circ$02$^{\prime} 02.3^{\prime \prime}$,
and the FOV is 5.2$'$. The image on the right is centered on E0102 and
the FOV is 2.7$'$. }
\label{E0102N76neconoiv}
\end{figure}

We compared the line and continuum maps of E0102 with those of the
H~II region, N76, to its southeast (see Figure
\ref{E0102N76neconoiv}), in order to examine their infrared colors and
to assess the effect of background contamination by N76.  E0102 shows
stronger \neiif\ and weaker \oivf\ emission than N76.
The Ne lines of the SNR are thus confidently
measured.  The continuum and \siiif\ emission of N76 is strongest at
its outer edge, while \oivf\ is strongest at the inner shell of N76,
as shown in Figure \ref{E0102N76neconoiv}. Since E0102 has its strongest
\neiif, \oivf, and dust continuum emission in the southern and eastern
parts, we are confident that the Ne, O and continuum measurements are
not contaminated by N76. 
The \siiif\ line image 
shows that the SNR also emits \siiif\ emission as well as the H~II
region. Therefore, the \siiif\ line in the IRS spectra likely belongs
to the SNR. We use the spectrum without local-background subtraction
in Figure \ref{E0102irsb} for the spectrum of E0102, but
the difference from the  local-background subtracted spectrum is used in 
estimating the uncertainty of the spectrum.

Infrared ejecta seem to most closely correlate with optical emission.
Figure \ref{nexrayhstcontours} demonstrates the correlation among
the X-ray, optical and three of our infrared ejecta maps.  Infrared
\oivf\ is a fair direct comparison to the optical [O~III] line maps,
but the \oivf\ image has a factor of few worse angular resolution than
that of the \neiif\ image.  \neiif\ and \oivf\ correlate with optical
[O~III] images in the southwestern shell and center. It is likely the
infrared ejecta are denser, cooler ejecta than the optical ejecta. In
the future, the infrared and optical data should be combined to more
accurately estimate of the ejecta mass, because a wide range of
ionization states are observed in the remnant. However, it will be
more meaningful to combine them within similar resolution images,
since the ejecta knots have very small scales not only in optical but
also in infrared as shown in the case of Cas A \citep{enn06}. The dust
continuum emission correlates with [O~III] and X-ray emission for most
of E0102 as well as with infrared ejecta in the southwestern and
central parts of the SNR.
This may indicate that the freshly formed
dust is correlated not only with infrared ejecta, but also with
optical and X-ray ejecta (recall that the infrared continuum shell is
still located inside the forward shock so it is less likely due to
CSM/ISM material).  Dust forms in the ejecta when the gas cools below a
temperature of 1000-2000 K. 
 After the reverse shock encountered the ejecta, it sufficiently heated both
gas and dust, which emit the infrared emission lines and continuum,
respectively.

\subsection{Doppler-shifted lines and maps}

\begin{figure}
\epsfig{figure=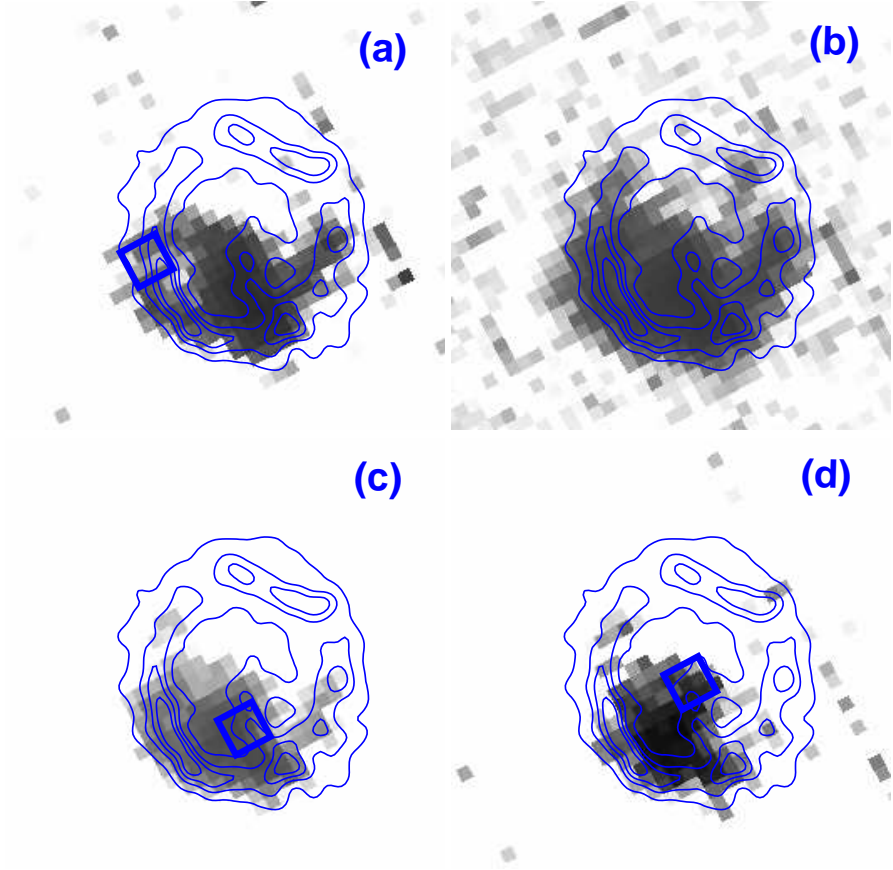,width=12truecm}
\caption{Line maps [Ne~II] (12.8$\mu$m) showing blue- to red-shifted
  wavelength regions: 12.5-12.7 \mic\ (a), 12.7-12.8 \mic\ (b),
  12.8-12.9 \mic\ (c), and 12.9-13.1 \mic\ (d).  The maps are
    overlaid with the Chandra X-ray contours.
    The positions of the spectra in Figure \ref{E0102irsbroaden} are
    marked as boxes in the maps, (a), (c) and (d) for the locations of "E", "Scenter",
    and "center", respectively (see Table \ref{tlinebroad}). }
\label{E0102dopplermaps}
\end{figure}

\begin{figure}
\plotone{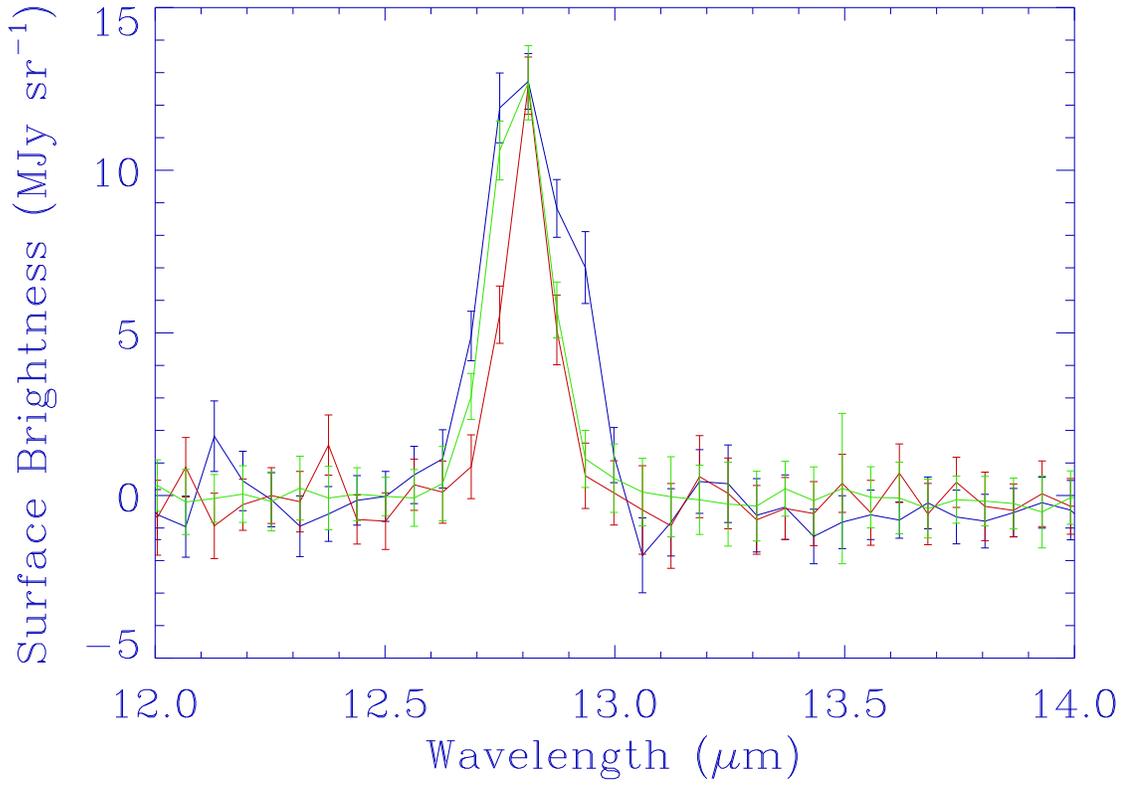}
\caption{Spectral line profiles of \neiif\ toward three positions (see Table \ref{tlinebroad})
 in E0102: Center (blue), E (green), and Scenter (red). The line
 broadening in the Center spectrum is consistent with a velocity of 5226 km s$^{-1}$ (see
 Table 3 for details).}
\label{E0102irsbroaden}
\end{figure}

The measured \neiif\ line width of the staring-mode spectrum (SE shell position in
Figure \ref{e0102irsslit}) is
0.1529$\pm$0.0023 \mic\ (Table \ref{tlinebroad}).
To examine the velocity broadening,
we extracted SL1 spectra from three additional positions in E0102--center, east, and Scenter 
(see the boxes of Figure \ref{E0102dopplermaps}).  
The three spectra show different line widths,
as shown in Figure \ref{E0102irsbroaden}.  
The doppler-shifted line maps in Figure \ref{E0102dopplermaps}, 
with the locations of the three spectra overlaid 
show that the central position has both blue- and red-shifted emission,  while
the eastern position has blue-shifted emission. 
The properties of the line
widths at these positions are summarized in Table \ref{tlinebroad}.  At
the center, the measured $\sim$0.12 \mic\ line width is comparable to
those measured for the \neiif\ calibration sources G333.9, NGC6720 and
NGC7293.  By contrast, the widths in the center and E spectra do
indicate line-broadening.

Fitting one Gaussian component, the line widths correspond to velocity
dispersions of 5226$\pm$304, 3579$\pm$54 km s$^{-1}$, and 3433$\pm$64 km s$^{-1}$ for the center,
SE shell, and
E, respectively. After taking into account the spectral resolution,
we can estimate the true velocity dispersion for each
position (Table \ref{tlinebroad}). The center, E and SE shell positions show the
true velocity dispersion of 4446$^{+353}_{-361}$, 2060$^{+104}_{-108}$,
and 2295$^{+85}_{-83}$ km s$^{-1}$, respectively. The observed velocity
shifts and dispersions are comparable to those from optical
measurements from -2600 to 3640 km s$^{-1}$ \citep{tuo83,bla00}.
The broad line profile for the center can also
be fitted with two Gaussian components.  Using a rest wavelength of
\neiif\ of 12.8135 \mic, the ejecta material must then be red-shifted by 2615
km s$^{-1}$ and blue-shifted by 641 km s$^{-1}$. 
 A confirmation with higher-spectral resolution data is desirable because
of the limited spectral resolution of IRS low-resolution spectra.

\subsection {Ne Line-Flux Ratio}

 In the line list in Table \ref{table:tlineflux}, the ions \neiiif\ and \nevf\ offer pairs
 of lines which are suitable for line diagnostics in order to
 constrain densities and temperatures.  The measured line flux ratios
 for [Ne V] $\lambda$ 14.3/24.3$\mu$m are 1.76$\pm$0.11, and for
 [Ne~III] $\lambda$ 15.6/36 $\mu$m 54$\pm$28. The ratio 15.6/36 $\mu$m is
 assigned an additional 50\% systematic error for the 36 $\mu$m line
 intensity, because this line falls on the degrading part of the
 array.  To constrain temperatures and densities, we calculate the
 line intensities and ratios of [Ne V] and [Ne III].  We solve the
 excitation-rate equations including collisional and radiative
 processes as a matrix using 5 energy levels for [Ne V]
and 3 levels for [Ne III].  The input atomic data were
 taken from \cite{gri00}\footnote{see
   http://www-cfadc.phy.ornl.gov/data$\_$and$\_$codes/aurost/aurost$\_$excit/}
 and include temperature-dependent collisional strengths. The
 solutions of the \nevf\ line diagnostics differ between the newer
 temperature-independent collisional strengths of \cite{gri00} and
 older values \citep[e.g.,][]{ost89}, while those of \neiiif\ are
 almost the same. 
 The line diagnostics are shown in Figure \ref{nediag}. The density
 and temperature jointly obtained from the ratios of \nevf\ and
 \neiiif\ are 7$\times$10$^4$ ($>$9000) cm$^{-3}$ and
 612$^{+300}_{-200}$ K, when assuming that both [Ne~III] and [Ne~V] gas
 come from the same gas.  Such a high density seems to be
 characteristic of SN ejecta as observed in optical lines;
 \cite{che79, che78}, for example, suggested that the optical \siif\ line in
 Cas~A originated from ejecta with a density of 10$^5$ cm$^{-3}$.

\section{Discussion}

\subsection{Dust Spectral Fitting and Dust Mass of E0102\label{DustSpecFit}}

The IRS spectra of E0102 have a prominent dust feature
peaking at 18 \mic\ which coincides spatially with the emission
from the infrared-emitting ejecta. This is clear evidence that dust
is forming in the ejecta. 
To determine the dust composition and mass, we performed spectral
fitting to the IRS dust continuum using the deep IRS spectra shown in
Figure \ref{E0102irs}.  
Note that the wavelength range
used for the fits is 5-40 \mic, and that only the continuum emission is
used to estimate the dust mass.

First, we estimated the contribution of the synchrotron emission to
the infrared spectrum using the radio fluxes and spectral index. The
radio flux at 408 MHz is 0.65 Jy for the entire SNR \citep{amy93}.  We
measured the radio surface brightness at the IRS position to be 0.58
MJy sr$^{-1}$ from the 408 MHz image of \cite{amy93}.  For a radio
spectral index $\alpha=-0.70$ \citep{amy93}, where 
$S_\nu\propto\nu^\alpha$, the expected synchrotron fluxes are
$1.4\times 10^{-4}$, $3.7\times 10^{-4}$, and $5.5\times 10^{-4}$ MJy
sr$^{-1}$ at 5, 20 and 35 \mic, respectively.  This is a very small
contribution to the infrared continuum ($<$1\%).

The dust continuum was then modeled with the Planck function
$B_\nu(T)$ multiplied by the absorption efficiency ($Q_{abs}$) for
various dust compositions, with the amplitude and temperature of
different components allowed to vary.  The fitting technique and mass
estimation method is the same as that was used for Cas A (Rho et al. 2008).
The spectrum shortward of 10 $\mu$m is highly uncertain due to background
subtraction and instrumental noise. 
Three dust components are needed to fit the IRS spectrum of E0102.

The 18 \mic-peak dust feature includes silicates, but it cannot be
reproduced by the combination of C, Mg-rich silicates, and Al$_2$O$_3$
grains that are most likely to be associated with the Ne- and O-rich
ejecta. 
A component with a feature around 16 $\mu$m is also needed.
Candidates include MgO and Si.  Whereas the
MgO feature calculated using the optical constants of
\citet{hofmeister03} turns out to be too narrow, amorphous Si, which can
condense in the C-rich layer in the primordial SNe II \citep{noz03},
can successfully reproduce the observed feature.

The rest of the continuum requires either carbon (Model A) or
Al$_2$O$_3$ (Model B) dust, as shown in Figures
~\ref{E0102dustfitcarbon} and ~\ref{E0102dustfital2o3}.
We favor
carbon or Al$_2$O$_3$ over solid Fe dust, because we expect carbon or
Al$_2$O$_3$ dust to be present where the Ne and O ejecta lines are
dominant: Ne, Mg, and Al are all carbon-burning nucleosynthesis
products \citep{woo95}.  MgSiO$_3$ and Al$_2$O$_3$ are condensed in
the ejecta of Ne, Mg, and Al, which are also carbon-burning products;
the O and Al are found in the outer layers of ejecta, which is where
Si and carbon dust grains condense. The dust and ejecta compositions
seen in E0102 are similar to those of the weak 21 $\mu$m dust layer in
Cas A \citep{rho08}. E0102, however, does not have the same silicate
dust features seen in Cas A.  Also, unlike Cas A, the dust composition
in different ejecta layers in E0102 cannot be spatially resolved, at
least partly because observations of E0102 suffer from limited angular
resolution due to its 18-times greater distance.

The dust compositions, temperatures, and masses obtained from the best
fits are summarized in Table ~\ref{tbestfit}. The estimated dust masses
within the IRS observed slit are 5.0$\times$10$^{-3}$ and
5.5$\times$10$^{-3}$ M$_{\odot}$ for Models A (carbon) and B
(Al$_2$O$_3$), respectively. To determine the sensitivity to dust
temperature, in Model C we performed the fit allowing the temperature
of Al$_2$O$_3$ to vary independently.  Model C produced a low
temperature ($<45$ K) that cannot be constrained by our spectra, which
lack long-wavelength data; therefore, we don't consider Model C further
and fixed the temperature of Al$_2$O$_3$ in
Model B to that of carbon in Model A. To explore the effect of
including many minerals, in Model D we performed a fit including all
the most common dust compositions (i.e., MgSiO$_3$, Si, Al$_2$O$_3$,
carbon, and Mg$_2$SiO$_4$; see e.g.T01, N03).  The mass is relatively
well constrained by the observations (within 20-40\%).  The fit of
Model D is shown in Figure \ref{dustcombination}. However, the reduced
$\chi^2\sim 2$ is still relatively high, largely due to the residuals
near where the gas lines were subtracted. (Note that some of lines
show kinematic broadening.)

 To estimate the total dust mass, we correct the observed
mass 
(within the high-sensitivity spectral slit)
 for the fraction of the SNR
covered by the IRS slit using a continuum map (between 21 and 23.5 \mic) generated from the
lower-sensitivity spectral cube.
This approach is justified 
because the spectra from the cube did not show significant variation across the SNR.
The correction corresponding to the total flux of the entire SNR
relative to the flux of the regions covered by the IRS slit is a
factor of 2.9.  The total dust masses for the entire SNR are then
0.015, 0.007, and 0.014 M$_{\odot}$ for Models A, B, and D,
respectively.  We favor Models A or D over Models B or C, because
the temperature of Al$_2$O$_3$ could not be constrained.

MgSiO$_3$,  Al$_2$O$_3$ and carbon are major dust species  predicted to
be produced in the ejecta of SNe by both  \cite{tod01,noz03}.  Note
that presolar Al$_2$O$_3$ grains from meteorites have been inferred to
be among the most abundant isotopically-enriched materials ejected by Type
II SNe \citep{cla04}. In terms of
amorphous Mg$_2$SiO$_4$ versus MgSiO$_3$, the young SNRs Cas A
\citep{rho08} and SN 2006jc \citep{noz08} also show a larger mass of
MgSiO$_3$ than  Mg$_2$SiO$_4$, whereas \cite{noz03} and  \cite{tod01} predict the
reverse. This may suggest that the chemical network favors MgSiO$_3$ 
over Mg$_2$SiO$_4$.

\subsection{ Ionization States of Ne}

\begin{figure}
\epsscale{1.05}
\epsfig{figure=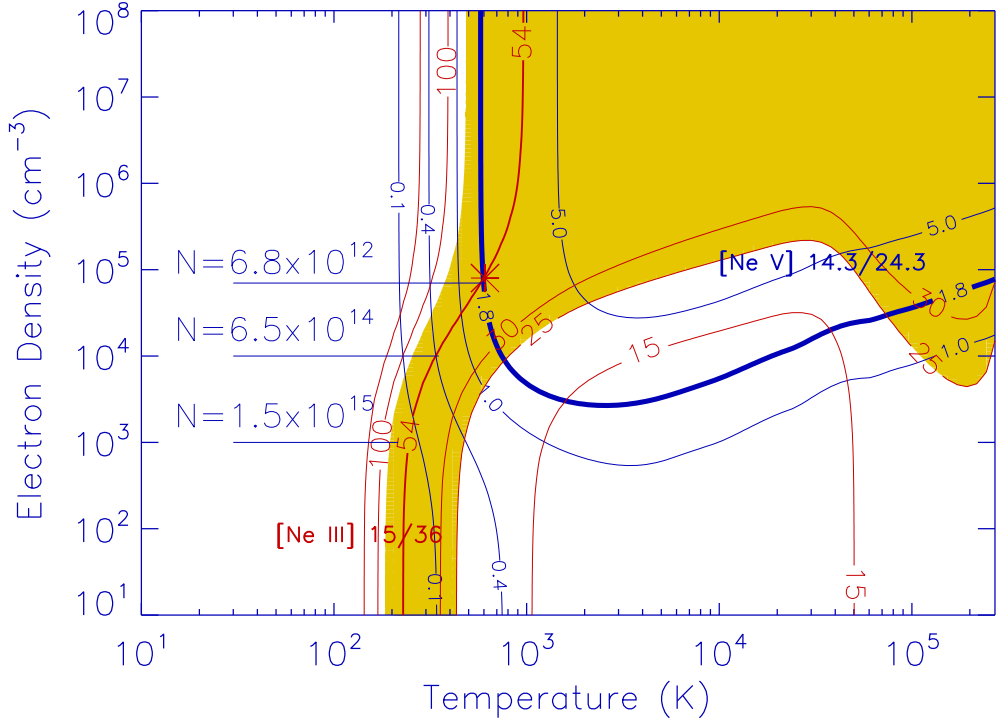,width=15truecm}
\caption{Line diagnostic contours of the ratios of [Ne~III] 15/35 \mic\
(red) and [Ne~V] 14.3/24.3 \mic\ (blue).  The observed ratios are marked
with thick solid lines for the best value of the ratios \neiiif\  = 54
and \nevf\ = 1.76. Note that the density is an electron density.
The shaded region shows the range of temperatures and densities
allowed by errors for the \neiiif\ ratio, whereas  the allowed physical
conditions of \nevf\ lines are marked with the thick blue contour lines.
The contours of [Ne~III] 15/36 and [Ne~V] 14.3/15.3 intersect at the
solution with a temperature of 612 K and a density of 4$\times$10$^{4}$
cm$^{-3}$ (marked as an asterisk).  The column density of [Ne~III] is
marked for three different densities (see Table  \ref{Tneionization}).
}
\label{nediag}
\end{figure}

\begin{figure}
\epsscale{0.7}
\plotone{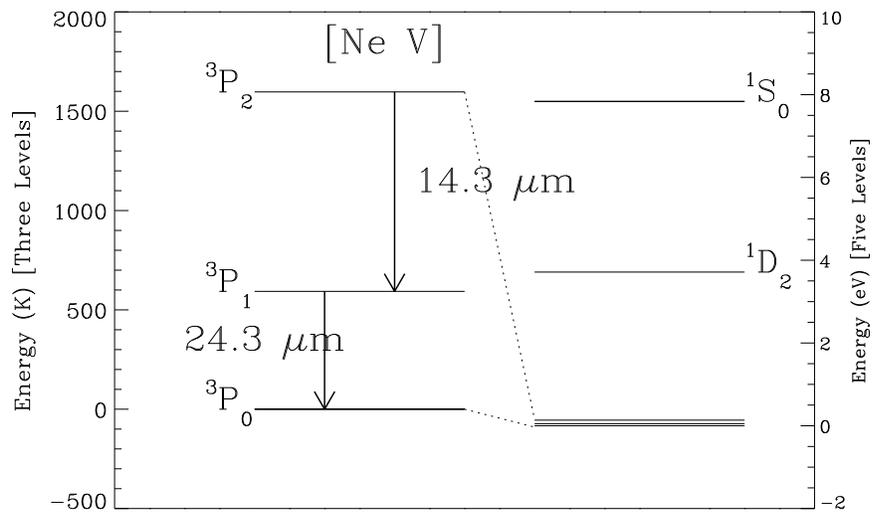}
\caption{Energy diagram of \nevf. Five levels are shown to the right
in a unit of eV, with a zoom-in of the three ground-stage levels to the
left (with a unit of K). }
\label{nevenergy}
\end{figure}

\begin{figure}
\epsfig{figure=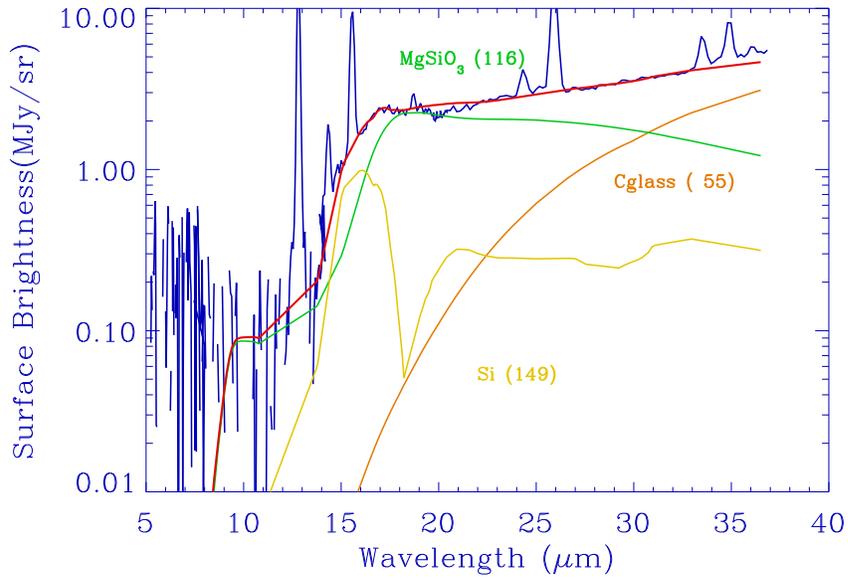,width=12truecm}
\caption{E0102 IRS dust spectrum superposed on a dust fit (Model A in
Table \ref{tbestfit}). 
 The continuum was fitted with dust compositions
of  Si, MgSiO$_3$ and Carbon.  The compositions suggest that the dust
forms around carbon-burning layers. The data and the total fit are shown
in blue and thick red lines, respectively. Contributions from each type
of dust are shown with the dust temperatures given in parentheses. The gas lines are 
also shown in the plot, but
they are excluded in spectral fitting
 (see Figure \ref{E0102dustfital2o3}) }
\label{E0102dustfitcarbon}
\end{figure}

\begin{figure}
\epsfig{figure=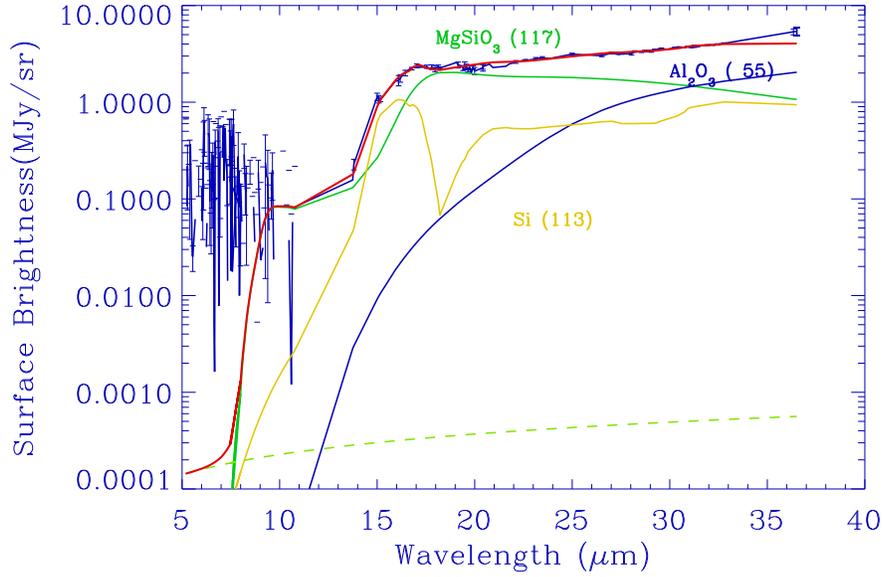,width=12truecm}
\caption{The dust continuum superposed on the dust fit of Model B
(Table ~\ref{tbestfit}) 
with the dust compositions of Si, MgSiO$_3$,
and Al$_2$O$_3$.  The synchrotron continuum contribution estimated
from the radio fluxes and spectral index is shown by the green dashed
line.
\label{E0102dustfital2o3}}
\end{figure}

\begin{figure}
\epsfig{figure=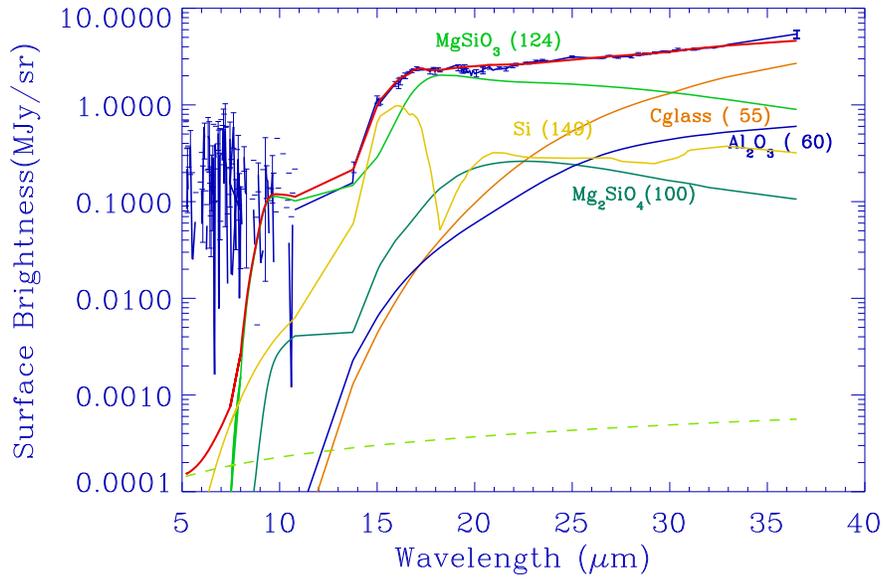,width=12truecm}
\caption{The dust continuum superposed on a combination of dust models 
of Si, MgSiO$_3$, Al$_2$O$_3$, Mg$_2$SiO$_4$, and carbon.
\label{dustcombination}}
\end{figure}

Column densities for [Ne III] and [Ne V] can be determined from
the excitation matrix given by: 
\begin{equation}
I(i, j) = \left({1\over{4 \pi}}\right)\, N_{i} \, A(i,j) \, h \nu_{ij} 
\end{equation}
where $I(i,j)$ is a line intensity between levels i and j, 
$N_i$ is a column density in the level i,
 $A (i,j)$ is spontaneous radiation transition rate,
$N_i$ is a column density in the level i, and  
$\nu_{ij}$ is the energy difference between i and j levels. 
For level $i$, the matix to solve the column density N$_i$ is given by:
\begin{equation}
\Sigma_{j<i} \, N_i\, A_{ij} + \Sigma_{j} \, N_i\, n \, \gamma_{ij}
= \Sigma_{j<i} \, N_j\, A_{ji} + \Sigma_{j} \,N_j\, n \, \gamma_{ji}
\end{equation}
\begin{equation}
 {{\gamma(j,i)}\over {\gamma(i,j)} } \,= {g_{i} \over {g_{j}}} \,
\exp^{-h \nu_{ij}/kT} 
\end{equation}
 where $n$ is density, $T$ is temperature and
$\gamma(i,j)$ and $\gamma(j,i)$ the collisional transition and
deexcitation rates.  The total column density of N = $\Sigma_i N_i$ is
obtained by solving the matrix. For example, the total column density
of [Ne~V] is the sum of all five levels as shown in Figure
\ref{nevenergy}.  For a temperature of 612 K and a density of
7$\times$10$^4$ cm$^{-3}$, the column densities for [Ne~III] and [Ne~V] are
6.8$\times$10$^{12}$ and 5.9$\times$10$^{11}$ cm$^{-2}$, respectively.
We estimated the column density for a few sets of densities and
temperatures implied by the observed \neiiif\ ratio as shown in Figure
\ref{nediag}, and found that the column density changes rapidly depending
on the electron density as listed in Table \ref{Tneionization}.  We
also solved the excitation-rate equation using 2 energy levels for Ne
II. The observed line brightness of \neiif\ can be used to estimate
the ion column density to be 2.7$\times$10$^{13}$ cm$^{-2}$.  The
column densities, ionization potentials, and critical densities of Ne
are summarized in Table \ref{Tneionization}.

The nucleosynthetic yields of Ne trace the progenitor mass.  In order
to obtain the Ne ejecta mass, we must estimate the column densities of
the remaining ionization stages of Ne.  First we obtain the column
densities when we assume local thermodynamic equilibrium, which is
described below in this section. Later we use shock models to estimate
the column densities of ionization states as described in
  \S4.3.
  
The Saha equation is given by:


\begin{equation}
\ln\left({{N_{r+1}}\over{N_r}} N_e\right) = \ln
\left[{U_{r+1} \over U_r} {{2 (2\pi m_p k
T)^{3/2}}\over h^3}  \exp \biggl({-{\chi}\over{kT}}\biggr)\right]  \\
\sim \ln\left({{U_{r+1} \over {U_r}}}\right) + 36.11413 + {3\over2}{\ln} T - 11602.81
{{I_{eV}} \over T}
\end{equation}
where  N is the number density of atoms in the r
(or r+1)th stage of ionization, and U is the partition function depending
on the temperature, $m_p$ is the mass of the proton, T is the temperature in
Kelvin, $\chi$ (I$_{eV}$) is the ionization potential for r to (r+1) (in units of eV).  

\begin{figure}
\epsfig{figure=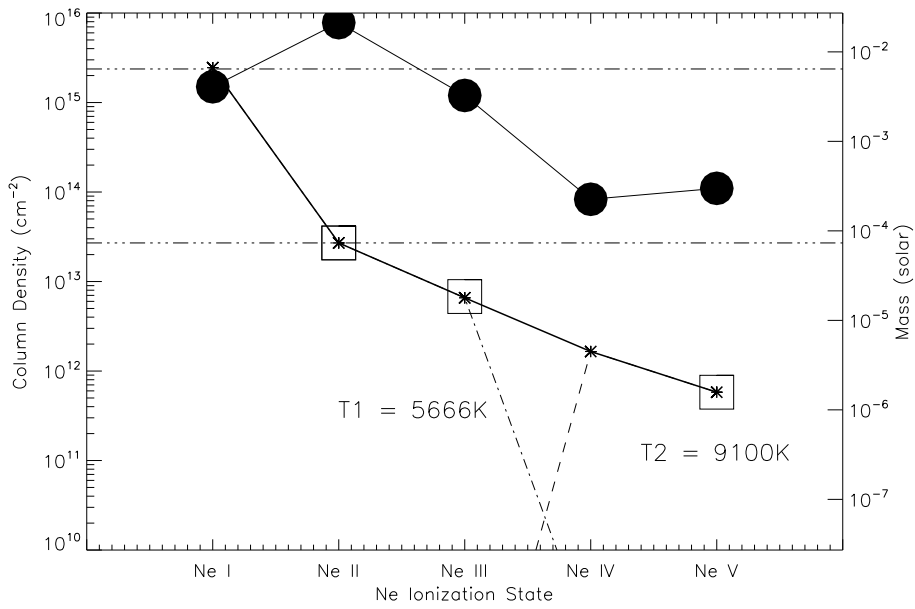,width=12truecm}
\caption{Column density for each Ne ionization state.  
The column densities are derived from collisional ionization  
(squares)
and from shock models (Filled dots).
The low (5667 K, dash-dotted line) and high (9300
K, dashed line) temperature components are marked with the sum (solid
line) of the two temperature components.  
The corresponding Ne mass is
shown in the right side of   y-axis.  The horizontal lines
(dash-three-dotted line) are [Ne~I] and [Ne~II] column densities and
masses.}
\label{neionization}
\end{figure}

We estimated the column densities of [Ne~I] and [Ne~IV] using the Saha
equation repeatedly as follows.  First, we obtained a gas temperature
of 5667 K by using the column densities of [Ne~II] and [Ne~III].  While
this is a much higher temperature than obtained from the [Ne~III] and [Ne~V]
line diagnostics, it is consistent with the error range for the line
ratios (see Figure \ref{nediag}).  Second, we estimated the neutral 
[Ne~I] column density by using the column density of [Ne~II] and a
temperature of 5667 K.  Third, we estimated the [Ne~V] column density by
using the temperature of 5667 K and the Saha equation, and found that
the inferred [Ne~V] column density was much smaller than the observed
value of 5.8$\times$10$^{11}$ cm$^{-2}$.  This indicates that most of
the observed column density of [Ne~V] comes from a gas with a
temperature different from 5667 K. An additional component with a
temperature of 9100 K is required to fit the observed [Ne~III] and [Ne~V]
column densities.  The [Ne~IV] column density is then obtained by
interpolating between [Ne~III] and [Ne~V].  Figure \ref{neionization}
shows the column density for each ion of Ne from the two-temperature
components derived here.  Column densities for the remaining, more
ionized Ne can be calculated for these two temperatures, and can be
seen to fall rapidly with increasing ionization.  We also estimated
the column density of [Ne~III] for (n$_e$, T) = (10$^4$ cm$^{-3}$, 350 K)
and (10$^3$ cm$^{-3}$, 215 K). We find that the column density varies
depending on density, as summarized in Table \ref{Tneionization} and Figure \ref{nediag},
and that it does not vary much with temperature.
A lower electron density yields a higher column density.

\subsection{Shock Model and Neon Mass}
We have calculated radiative shock models for shocks in the velocity range of
50 - 500 km s$^{-1}$.  These models use many of the same assumptions used by
\citet{bla00}. In particular we use the same elemental abundances ($\mathrm{H}
= 12.00, \mathrm{O} = 16.00, \mathrm{Ne} = 15.50, \mathrm{C} = 14.50,
\mathrm{Mg} = 13.90$) and follow their approach of using a single ram
pressure, $n_0 v_s^2 = 16\,\mathrm{cm}^{-3} 100\,\mathrm{km}\,
\mathrm{s}^{-1}$, for each shock speed in the standard case.  We also follow
\citet{bla00} in using a magnetic field strength of $B_0 = 4 \sqrt{n_0}\,
\mu$G for our standard case.  \citet{bla00} use the results of a range of
shock models with shock velocities ranging from $30 - 500$ km s$^{-1}$,
weighting the emission by the fraction of the shock area that is assigned (by
assumption) to each shock speed.  We take the simpler approach of finding the
best fitting single shock speed.  In addition to the standard cases, we have
calculated models with ram pressures that are larger by a factor of 1.5, 2 and
3 and with smaller magnetic field strengths, down by a factor of 2.  The model
calculations assume 1-D (plane parallel) steady flow and include
non-equilibrium ionization, radiative cooling and magnetic pressure, $B
\propto n$.  We also include photoionization from radiation generated in
upstream gas and the accompanying heating.  Since the region of strong
ionizing radiation emission is mainly in the hot, ionized portion of the
shock, while the region of substantial photoionization is downstream, this
approach is satisfactory.  In fact photoionization and heating are negligable
within the X-ray/EUV emitting zones where $T > 10^4$ K. These calculations make
use of a revised version of the \citet{RS_77} plasma emission code for the
calculation of the hot, ionized post-shock zone and the cooling zone in which
the gas temperature drops sharply to $\lesssim 1000$ K.  

\begin{figure}
\epsfig{figure=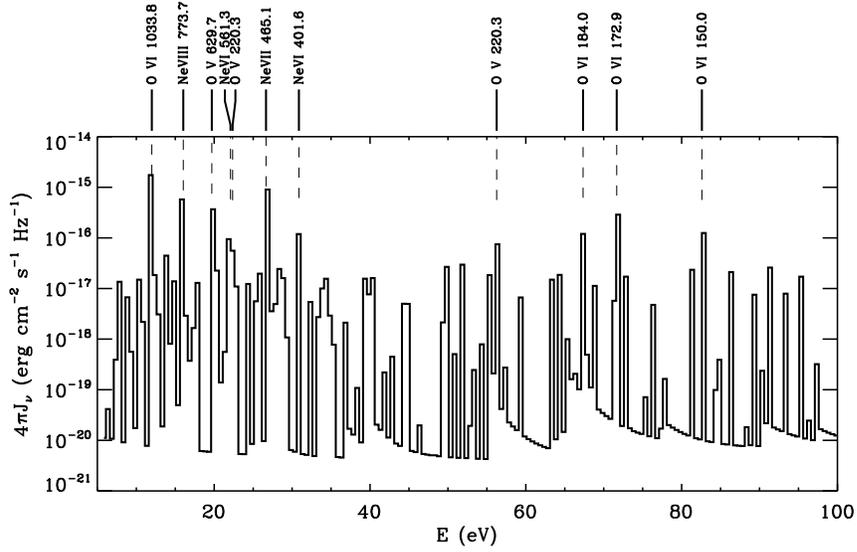,width=12truecm}
\caption{Ionizing radiation generated in the shock and incident on the gas in
the photoionization zone for the case V$_s \sim$200 km s$^{-1}$ and $3\times$
the \citet{bla00} ram pressure.  The spectrum is clearly line dominated and
many of the lines are from O and Ne which have highly enhanced abundances in
the ejecta.}
\label{fig:radfield}
\end{figure}

\begin{figure}
\epsfig{figure=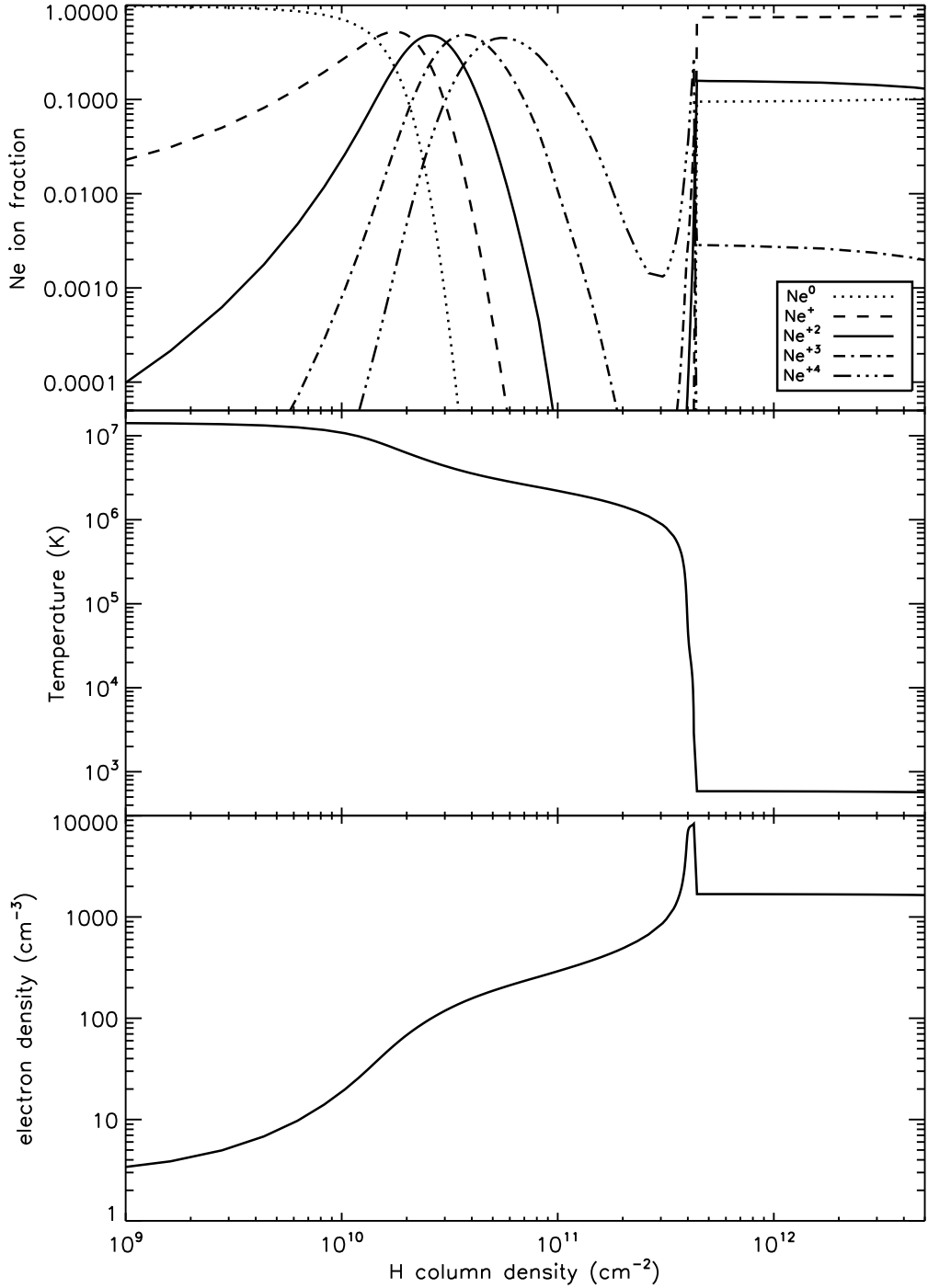,width=14truecm}
\caption{Neon ionization (above) and gas temperature (below) in the shock for
our model with shock speed 200 km s$^{-1}$ and ram pressure 3$\times$ the
\citet{bla00} assumed value.  The ionization of Ne clearly shows the lag in
ionization and recombination in the ionization ($T \geq 10^5$ K) and
cooling ($10^3 <\leq T \leq 10^5$ K) zones.  The region on the right,
where $T \sim 550$ K, is photoionized and assumed to be in thermal and
ionization equilibrium.  The Ne$^{+4}$ ion fraction is very low in the
photoionized zone and so the observed [\ion{Ne}{5}] 14.3 $\mu$m and 24.3
$\mu$m emission come from the cooling zone (an electron density of
$10^3-10^4$cm$^{-3}$) whereas most of the [\ion{Ne}{2}] and [\ion{Ne}{3}]
emission comes from the photoionized zone.}
\label{fig:Ne_ioniz}
\end{figure}

For the calculation
of the photoionized zone, we use the Cloudy code \citep[version 07.02.02 of
the code last described by][]{Ferland_etal_1998} 
with the incident radiation field
taken to be the ionizing radiation field generated in the hot,
collisionally ionized and cooling zones (as calculated using the R\&S code).
We show an example of the ionizing radiation field in Figure \ref{fig:radfield}.
Cloudy assumes thermal and ionization equilibrium (in contrast to our
calculations using the R\&S code), but this should not be a very bad
assumption for the cool ($T\sim400 - 1000$ K), dense gas of the photoionized
zone. The intensity of the ionizing radiation field depends on the geometrical
factor, $R_\mathrm{max}$, related to the ratio of the lateral extent of the
shock front to its thickness \citep[see, e.g.][]{Raymond_1979}. We have
generally used $R_\mathrm{max} = 1$ in our calculations, but found the best
agreement for a model with $R_\mathrm{max} = 0.7$ indicating small shock
fronts as might be indicative of shocks propagating into clumps elongated in the radial
direction.  Figure \ref{fig:Ne_ioniz} illustrates the Ne ionization and
temperature for the two parts of the calculation for one model.

As illustrated in the figure, [\ion{Ne}{5}] emission is entirely from the
ionization and cooling zone, while [\ion{Ne}{2}] and [\ion{Ne}{3}] emission
comes primarily from the low temperature, $T \sim 600$ K, photoionization
zone. The electron density and temperature derived in Figure \ref{nediag} from
the intersection of the \neiiif\ and \nevf\ contours implicitly assumes a
uniform temperature and density for the \neiiif\ and \nevf\ emitting gas.
Since the emission comes from different regions with different temperatures
and electron densities, these diagnostics do not apply to emission from the
shock.

Under the assumption that the \neiiif\ and \nevf\ emission comes from the same
gas, we calculate the column densities of \ion{Ne}{1}, \ion{Ne}{2} and
\ion{Ne}{3} to be 7$\times$10$^{14}$, 3.5$\times$10$^{15}$,
6.5$\times$10$^{14}$ cm$^{-2}$, respectively, and the total Ne column density
to be 7.9$\times$10$^{16}$  cm$^{-2}$. To use the results of our shock models we
need to limit the extent of the photoionized zone.  We use the total \neiif\ 12.8
\micron\ emission to set the depth of the photoionized zone.  The total \nevf\
emission is naturally limited by the size of the post-shock ionization and
cooling zones.  We find, however, that we are unable to match both the total
surface brightness of \nevf\ 14.3 \micron\ emission and the ratio of 14.3
\micron/24.3 \micron\ emission at the same time when the surface brightness is
determined as in Table \ref{table:tlineflux}.  Matching the line ratio requires
lower shock speeds which in turn leads to more flux.  

This problem can be
overcome if the emission is clumped on scales unresolved by {\it Spitzer}.  Table
\ref{table:tlineflux} uses the default angular extraction size
(10\arcsec$\times$20\arcsec\ for LL) to estimate the surface brightness,
because the SNR structures were not resolved. However, we know from the other
young SNRs that the angular size of ejecta emitting regions are generally much
smaller than the implied $\sim 3$ pc.  Therefore, we estimate the angular size
(to get the solid angle) using a Ne image of Cas A \citep{rho08}.
From this we estimate the filling factor of the Ne emission over the
20\arcsec$\times$10\arcsec\ field of view to be $\sim 0.13$ in surface area.
The \nevf\ 14.3 \micron\ surface brightness of 3.974$\times$10$^{-6}$ erg
s$^{-1}$ cm$^{-2}$ sr$^{-1}$ thus becomes 3.057($\pm$1)$\times$10$^{-5}$ erg
s$^{-1}$ cm$^{-2}$ sr$^{-1}$ after the correction, where the errors are
estimated using experiments with different surface brightness contours (the
filling factor ranging from 0.1 to 0.2).  

\begin{figure}
\epsfig{figure=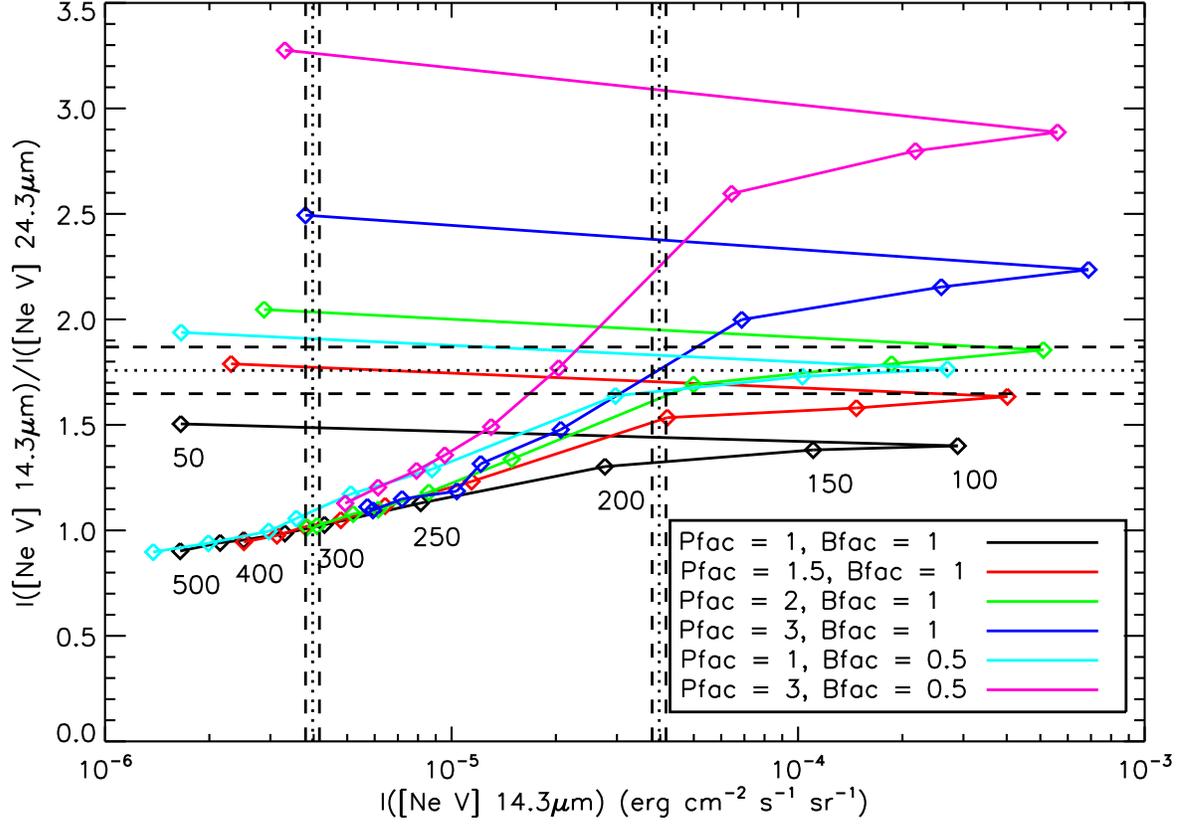,width=16.5truecm}
\caption{Predicted \nevf\ line brightness and the ratio of 14.3/24.3 \mic\ lines
depending on shock model parameters of shock velocity (indicated as numbers
for the standard case runs), ram pressure (P) and magnetific fields (B).  Pfac
and Bfac indicate the ratio of the value relative to the \citet{bla00} values
for the ram pressure and magnetic field.  The observed \ion{Ne}{5} lines are
best fit by two shock models of  ($Pfac=3$, $Bfac=1$) (blue line) and
($Pfac=1$, $Bfac=0.5$) (light blue line) with a shock velocity of $200-300$ km
s$^{-1}$.  The horizontal dotted line indicates the observed emission line
ratio and the dashed lines show the formal $1-\sigma$ errors. The leftmost
vertical lines indicate the 14.3 \micron\ surface brightness assuming that
emission fills the field of view.  If the unresolved emission regions only
have a filling factor of 0.13 then the true surface brightness is that
indicated by the vertical lines to the right.  Our ``best'' model assumes that
is the case.}
\label{fig:Ne_Vlines}
\end{figure}

With the correction for the filling factor we find that we can match both the
\nevf\ line ratio and fluxes for shock velocities of $\sim 200$ km s$^{-1}$,
though a better match is achieved for models with a higher, $\sim$ a factor of
3, ram pressure than the standard \citet{bla00} value, as shown in Figure
\ref{fig:Ne_Vlines}.  Using the filling factor correction and the best fitting
model for the \neiif\ and \nevf\ emission, we find the column densities of
[\ion{Ne}{1}], [\ion{Ne}{2}] and [\ion{Ne}{3}] to be 1.5$\times$10$^{15}$,
7.8$\times$10$^{15}$, and 1.2$\times$10$^{15}$ cm$^{-2}$, respectively, and
the total Ne column density of 4.0$\times$10$^{16}$ cm$^{-2}$. The column
densities from different ionization states of Ne are summarized in Table
\ref{Tneionization}.

\subsection{Ne Ejecta Mass}

From the column densities for each ion, we can readily calculate the
Ne mass by summing over all the ions as:
\begin{equation}
M = \Sigma_i N_i\, m_\mathrm{Ne}\, d^2\, \Omega
\end{equation}
where M is the total mass mass, $N_i$ is the column density of
ion $i$, $m_{Ne}$ is the mass of Ne, d is the distance and $\Omega$ the
emitting area.  In our calculation, we used a wavelength dependent $\Omega$
which is equivalent to the line extraction width (the approximate area of the
LL extraction region is 10\arcsec$\times$20\arcsec). 
The mass corresponding to
the column densities for each ion is also shown in Figure \ref{neionization}.
The total derived Ne mass (within the IRS slit) inferred from excitation and Saha equations  
is 0.22 M$_{\odot}$ for n$_e$= 7$\times$10$^4$ cm$^{-3}$ and T= 612 K,
and 0.07  M$_{\odot}$ for n$_e$= 1$\times$10$^3$ cm$^{-3}$ and T= 215 K, respectively. 
For the shock model we find a total mass in Ne of $0.014$ M$_\odot$ which 
supercedes the derived values.

This is the mass enclosed by the IRS slit only, which is just a
portion of the entire SNR.  We make the correction for the entire SNR
using the \neiiif\ map at 15.5\mic.  The total flux is 3.03 times
larger than that covered by the slit, so we obtain a total Ne mass 
(for all ions including neutral Ne) in
E0102 of $0.042$ M$_{\odot}$ 
from the shock model.

We compared our derived Ne mass of E0102 with that of SN2005af at an
epoch 214 days after the explosion as a rough consistency check. The Ne mass implied from
\neiif\ line of SN2005af is 2.2$\times$10$^{-3}$ M$_{\odot}$
\citep{kot06}. We determine the [Ne~II] mass in E0102 to be
3$\times$10$^{-3}$ M$_{\odot}$ (from the shock model), which is within a factor of 1.5 of that
determined for SN2005af, also a massive Type II SN. This is a
relatively small disparity given that there may be differences in the
progenitor mass, the respective physical conditions of Ne-emitting
gas, and thermodynamic history of Ne gas,
and also does not account for hotter and more ionized O ejecta that
are observed at shorter wavelengths.

We detected no infrared H lines from E0102, though a few do fall in
the IRS spectral band.  H lines are present in the optical spectra of
E0102, but it is not likely that they are associated with the
supernova ejecta \citep{bla00}.  For the purpose of estimating the Ne
abundance relative to solar (or cosmic) values, however, we adopt the
H$\alpha$ line flux 5.94$\times$10$^{-16}$ erg s$^{-1}$ cm$^{-2}$
and the upper limit of  H$\beta$ line flux of 1$\times$10$^{-16}$ erg s$^{-1}$ cm$^{-2}$ 
reported by \citet{bla00} in order to obtain an upper limit on the H column
density.  
At large
optical depth, 4~$\pi$~j(H$\beta$)/(N$_p$ N$_e$) is a constant for a
given temperature that varies by only a factor of 5 over the relevant
temperature range \citep{ost89}.
Taking the path length to be the aperture beam size of 1$''$ (0.299
pc), the H column N is inferred to be $<$7$\times$10$^{13}$ cm$^{-2}$.
The Ne
abundance in E0102 ([Ne/H]) using the Ne column density of
4$\times$10$^{16}$ cm$^{-2}$ is then $>$1.9$\times 10^{-3}$,
while the solar value ([Ne/H]$_{\odot}$) is 1.2$\times 10^{-4}$, so
[Ne/H]/ [Ne/H]$_{\odot}$ $>$18 if the optical H$\alpha$ is associated
with E0102.  More likely, that is not the case, so this is a lower
limit for the abundance of Ne relative to solar values.


We also calculated the column density of [O IV] using 5 levels for [O IV]
lines and the atomic data of \cite{tay06}. The column density inferred
from the observed \oivf\ line brightness is 1.5$\times$10$^{13}$
cm$^{-2}$. The equivalent oxygen mass is 8$\times10^{-5}$
M$_{\odot}$. We also estimated the column densities and masses of [O~I],
[O~II], [O~III], and [O~V] using the derived column density of [O~IV], the
Saha equation, and assuming that the oxygen gas has the same
temperature as the Ne gas.  The total mass of oxygen in the IRS
aperture is thus inferred to be 0.26 M$_{\odot}$. The calculated mass
is quite sensitive to the second component temperature, however.  For
example, a temperature of 7000 K rather than 9100 K reduces the O mass
a factor of two. The uncertainty in the estimated oxygen mass may thus
be a factor of a few.  To obtain the total oxygen mass for the entire
remnant we scale the masses above by the same 3.30 multiplicative
aperture correction factor used for Ne.  The total oxygen mass
estimated for the entire remnant is then approximately 0.78
M$_{\odot}$, as shown in Table \ref{yields}.
Note that oxygen mass estimate is much less accurate than that of
Ne because we only have one observed oxygen line.  

\subsection{Ejecta Mass and Nucleosynthesis Yields  }

It is evident that the infrared census of the Ne and O ejecta mass is
incomplete because the ejecta that emit at optical and X-ray
wavelengths are more ionized and at higher temperatures.  Infrared
emission traces dense and relatively low temperature ejecta as
indicated by our line diagnostics, whereas X-rays trace much hotter
and lower density coronal plasmas. 
The total neon and oxygen masses that we estimate from the infrared
observations and shock models are compared
to the nucleosynthesis yields of massive stars in \cite{thi96},
\cite{nom97}, and Chieffi \& Limongi (2003) as summarized in Table
\ref{yields}. 

The He- and H-like ions that dominate the X-ray emission are inferred
to have masses of 2 M$_\odot$ for Ne and 6 M$_\odot$ for O in the {\it
Chandra} study of \cite{fla04}; these indicate a progenitor mass of
32 M$_{\odot}$.  These mass estimates, however, are subject to
uncertainties in the composition of the underlying continuum.
\cite{fla04} make their mass estimate assuming that the ejecta do not
include light elements, but if this is not the case, \citet{fla04}
note that the X-ray derived ejecta mass could be a factor of 20 lower,
at 0.1 and 0.3 M$_{\odot}$, for Ne and O, respectively.  By comparison,
our estimates for the Ne and O ejecta masses from the infrared
observations are 0.042 M$_{\odot}$ and 0.78 M$_{\odot}$, respectively.
In the optical/UV analysis of \cite{bla00}, relative line strengths
and shock models favor a progenitor of 25 M$_\odot$, but no attempt is
made to estimate ejecta masses for the entire remnant.

\begin{figure}
\hbox{
\epsfig{figure=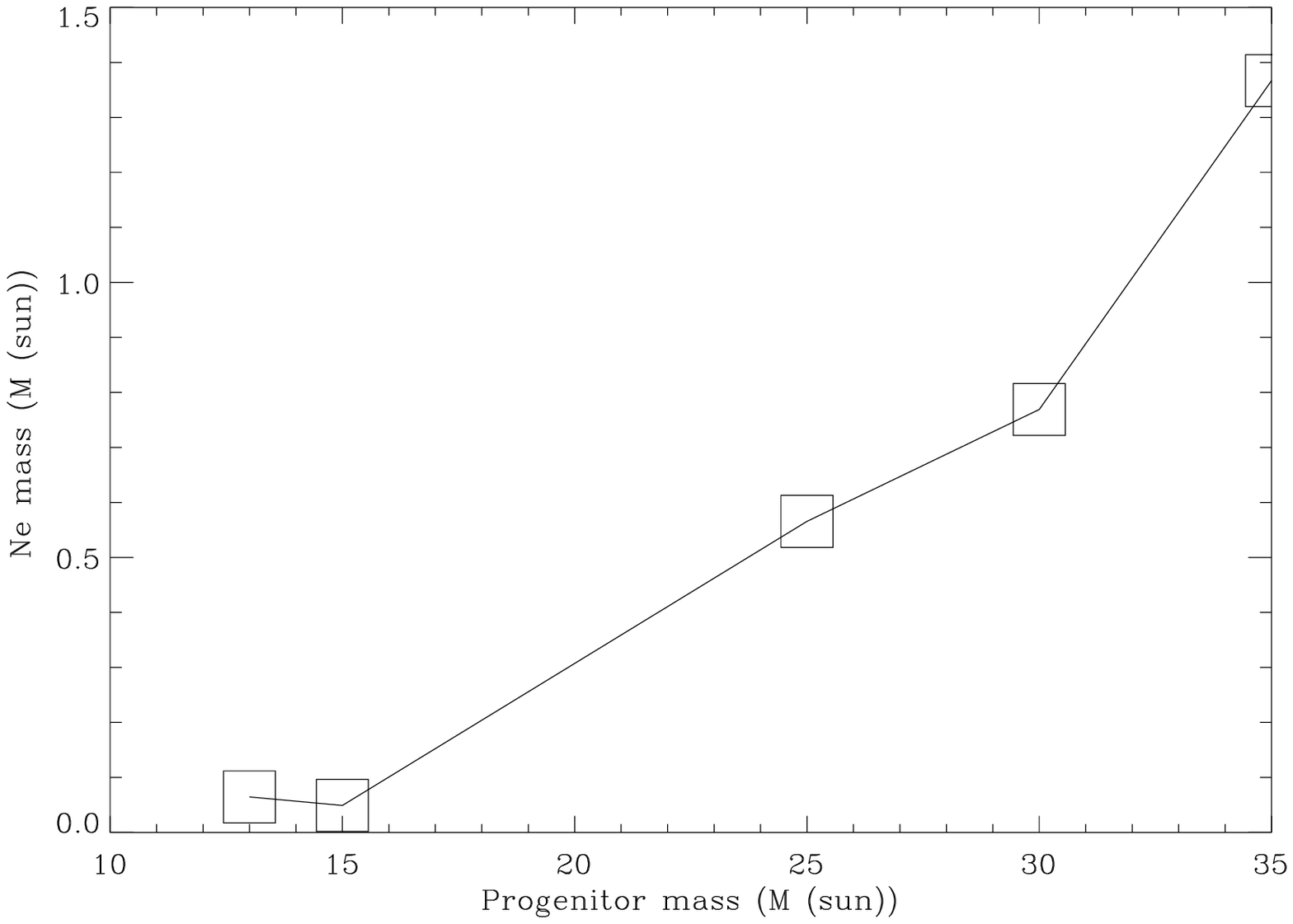,width=8truecm}
\epsfig{figure=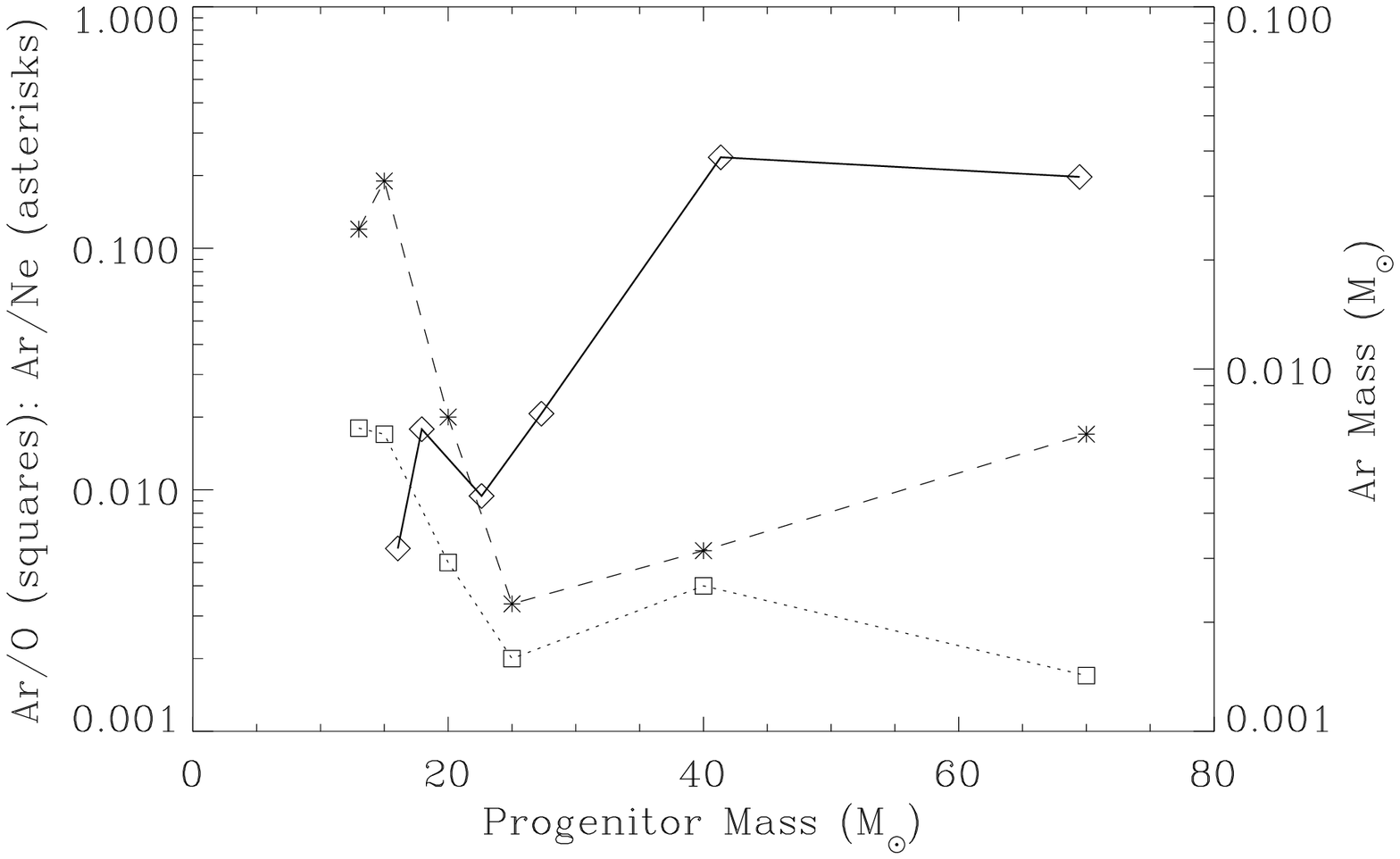,width=8truecm}
}
\caption{(a) Nucleosynthesis yields of Ne mass depending on the progenitor mass. (b)
Nucleosynthesis yields of Ar/Ne (dashed line), Ar/O (dotted line) ratios
and Ar mass (solid line) depending on the progenitor mass.}
\label{armass}
\end{figure}

If we take the lower set of values for the X-ray emitting ejecta
masses, we determine the total ejecta mass determined from the
combined infrared and X-ray analysis to be $>$0.142 M$_\odot$ for Ne
and $>$1.08 M$_{\odot}$ for O.  Taking the higher X-ray values, these
masses are 2.04 M$_\odot$ for Ne and 6.8 M$_\odot$ for O.  As a
comparison, the X-ray emitting Ne mass predicted by our shock model is
$\sim$0.84 M$_{\odot}$, though we note that this model was matched to
the infrared Ne line emission only.  Taken together, the inferred
ejecta masses favor a progenitor mass of 30 (25 - 40) M$_{\odot}$
as shown in Figure \ref{armass}a and Table \ref{yields}, but are
not strongly constraining.  The infrared Ne/O mass ($>$0.13) ratio by itself
favors a progenitor mass greater than 25 M$_\odot$, as do the total
masses of Ne and O when we take the lower range of the X-ray emitting ejecta mass
corresponding to the presence of light elements in the ejecta.  This
is not surprising given that the infrared ejecta mass makes up a
significant fraction of the total mass (30\% for Ne, and 70\% for O)
if the X-ray ejecta masses are taken to be at this lower limit.
Contrariwise, the infrared emitting masses make up a very low fraction
of the total ejecta mass (2\% for Ne and 12\% of O) if we take the
higher X-ray ejecta masses corresponding to pure heavy elements, and
the inferred progenitor mass is then necessarily in agreement with
that obtained in the X-ray analysis by \cite{fla04}.

The set of line-emitting ejecta elements is very different for E0102
and Cas A, although both are young core-collapse SNRs with optical
emission from high-velocity O ejecta.  In E0102, the Ar lines are
apparently absent in the infrared, and the Si and S lines are weak.
Cas A, by contrast, shows very strong Ar emission as well as strong
silicate (21 \mic-peak) dust emission (see the comparison in Table
\ref{tcompe0102casa}).  In addition, Cas A shows unshocked, centrally
located Si and S ejecta, whereas E0102 shows no indication of any
unshocked ejecta.  The differences in the X-ray line emission from
these two remnants are in line with those seen in the infrared.  E0102
is dominated by O and Ne \citep{fla04, sas06}, while Cas A is
dominated by Si, S, Ar, Ca, and Fe. The ejecta temperatures in Cas A
are higher, and presumably the O is too ionized to emit appreciable
line emission, although Cas A is also subject to much higher
interstellar absorption.  And in spite of being considerably younger
than E0102 (at an age of 330 yr), Cas A has a reverse shock that is
already traversing the innermost ejecta layers.  The key to this
difference may be the extensive mass loss experienced by the Cas A
progenitor, which probably had a binary companion (Young et
al. 2006).  Cas A is likely to have exploded at only 4 M$_\odot$ after
having started at $\sim$25 M$_\odot$ on the main sequence (e.g.,
Laming \& Hwang 2003).  This difference in the mass loss history of
the progenitors would also affect the supernova nucleosynthesis and
subsequent evolution of the remnant.

It is thus possible that the explosion of E0102 simply did not produce
a significant mass of Ar.  Certainly, variations of factors of 5-20 in
Ar mass are evident in various supernovae (for example, see the
comparison of SN 2005af to SN 1987A and SN2004dj in \cite{kot06}).
Nucleosynthesis models likewise predict a wide range of Ar masses,
with more massive progenitors producing significantly more Ar. The
yields range from 0.001 to 0.07 M$_{\odot}$ for 25-40 M$_{\odot}$
progenitors, with corresponding Ar/Ne mass ratios of 0.001 -- 0.1
(Figure \ref{armass}).  If we scale the predicted Ar/Ne mass
ratios to the total infrared Ne ejecta mass of 6.5$\times 10^{-3}$
M$_{\odot}$, the models predict an infrared-emitting Ar mass between
6.5$\times 10^{-6}$ to 6.5$\times 10^{-4}$ M$_{\odot}$. 

We can use the detection limit of 3$\times$10$^{-8}$ erg s$^{-1}$
cm$^{-2}$ sr$^{-1}$ for \ariif\ at 6.985 \mic\ to estimate the
corresponding mass of Ar in E0102.  Using two excitation levels of Ar,
the column density of Ar is 1$\times$10$^{10} $ cm$^{-2}$ if we assume
that Ar is at the same temperatures as Ne.  Since it is likely that the
temperature and density could actually be quite different from those
values, we consider a range of temperatures from 600 to 20,000 K and
densities from 10$^3$ cm$^{-2}$ to 10$^5$ cm$^{-2}$.  Including a
factor of 3 correction for the IRS aperture, the final upper limit for
the Ar mass ranges from 1.5$\times$10$^{-8}$ to 1$\times$$10^{-5}$
M$_{\odot}$.  The upper end of this mass range overlaps the range of
the expected infrared-emitting Ar mass estimated above.  It is thus
plausible that the current non-detection of Ar is simply be due to limited
sensitivity.

We cannot yet exclude the possibility that the infrared-emitting Ar
are truly absent, however.  The physical conditions of the
infrared-emitting Ar are highly uncertain.  Possibly, the reverse
shock has simply not reached the deep layers containing Si, S, Ar, and
Ca ejecta.  It is also possible that the Ar might be detected in
a more complete set of deep observations.  While the IRS mapping data 
are too
shallow to detect faint lines, the existing deep IRS staring mode
spectrum covered only one-third of the remnant. Given that
core-collapse remnants often show rather asymmetric ejecta
distributions, it cannot be ruled out that deep observations at other
positions in E0102 might reveal Ar ejecta.

\subsection {[Ne~V] lines: Comparison with Other Astronomical Objects}

Among various astronomical objects,  highly ionized \nevf\ is not
commonly detected in the infrared because of its high ionization potential
(97.11 eV),
whereas lines of both \neiif\ and \neiiif\ are common and often bright.
In galaxies, the detection of \nevf\ 24.3 $\mu$m line of comparable 
brightness to [Ne~II]
is often taken as a signature of active
galactic nuclei \citep{armus07} because \nevf\ cannot be readily
produced by O stars.
In the Galaxy, planetary nebulae emit
\nevf\ lines which are produced by white dwarfs with  high temperatures
\citep{rubin04, vanhoof00, rubin97}. E0102 is the first SNR which is
identified as an infrared \nevf\ emitter.

Here, we compare the physical conditions of ionic line emitting material
such as \nevf\ in the ejecta of E0102 with those of other astronomical
objects.  The [Ne V] 14.3/24.3$\mu$m ratio of 1.76 is similar to many 
planetary nebulae \citep{rubin04}, an example of which is NGC 7027. 
E0102 falls at the extreme high end of the range (0.2 to 1.8) observed
in active galactic nuclei \citep{dud07,sturm02}.  Both our (see Figure \ref{nediag}) and Dudik et
al.'s  calculations use the same collisional strengths from
\cite{gri00}, and we are able to reproduce the [Ne V] 14.3/24.3 $\mu$m
ratio plot \cite[Figure 1 of][]{dud07}.  Our measurements of \nevf\
ratios are not affected by the extinction correction, because the
intrinsic extinction correction is very low at less than 0.1\% for
E0102.  In the SNR E0102, \nevf\ lines come largely from collisionally
ionized and radiatively cooling zones, whereas \neiif\ is dominantly
from the photoionized zone. In active galactic nuclei, both \nevf\ and
\neiif\ lines are shown to be from photoionization.  The ratio of
\nevf/\neiif\ is only 0.06 for E0102 and about 1 for AGN.

\subsection{Dust Mass and Implication of Supernova Dust in Early Universe}
Our dust masses for E0102 are 8-18 times greater than the
$8\times10^{-4} M_\odot$ mass estimated by Stanimirovic et al.  (2006)
from MIPS imaging, but the latter assumes a temperature of 120 K and
typical interstellar dust composition.  The inferred dust mass will
depend on the composition and temperature of the dust, which can only
be accurately identified with spectral data such as the IRS data used
here.  The difference between our dust mass for E0102 and that of
Stanimirovic et al. (2006, hereafter S06) thus arises from a number of
factors.  First, the IRS spectra indicate that there are three
composition and temperature components.  While two of the three
components have temperatures comparable to the 120 K assumed by S06,
the third component had a much lower temperature (55-60 K), and as
shown in Table ~\ref{tbestfit}, dominates the mass.  Moreover,
because of the differences in dust composition assumed, there are
differences in the absorption coefficients used and also in the
inferred density; the latter are generally higher for the grains we
used here, with MgSiO$_3$, and Al$_2$O$_3$ having solid-state
densities of 3.2 and 2.45 g~cm$^{-3}$, respectively, compared to more
typical ISM grains, e.g., graphite with $\rho$ = 2.25 g cm$^{-3}$.

Our estimated dust mass is also higher than those generally inferred
for supernovae. 
Possibly, it is easier to
detect dust in young SNRs than in extragalactic SNe simply because the
dust in SNe is too cold to be detected in the infrared by {\it
  Spitzer}, whereas the freshly synthesized dust in young SNRs has
been heated by the reverse shock to observable temperatures.  Another
limitation for estimating the dust mass of extragalactic SNe is that the
spatial resolution currently available with IR telescopes for
observing supernovae is still poor, and that makes it difficult to
distinguish between fresh dust formed in the ejecta and ISM/CSM dust.
In the case of E0102, we were able to resolve and separate the ejecta
and the forward shocked materials, and found no significant infrared emission
from the ISM heated dust.

The estimated total dust mass of 0.007 to 0.015 M$_{\odot}$ for E0102
is a factor of 3 lower than the mass (from 0.02 to 0.054 M$_{\odot}$)
of Cas A \citep{rho08}, though of the same order of magnitude.  We
discuss a few possibilities for the discrepancy between our estimated dust
mass from E0102 and the dust mass per SN predicted by models
\citep[e.g.][]{tod01, noz03} to account for the dust in high
red-shifted galaxies.  First, there may be more mass in dust at lower
temperatures.  Our wavelength coverage is only up to 40 $\mu$m, and the
MIPS 70 \mic\ emission has not been detected due to confusion with the
nearby H~II region of N76 and the limited spatial resolution of the
70\mic\ image, as shown by \cite{sta06}. We could not constrain a
temperature of 41 K with our IRS data, as shown in Model C of Table
\ref{tbestfit}. Even if colder dust ($<<$55 K) exists, our observations
would not be sensitive to it. 
The colder dust components have been detected in Cas A and Kepler
(Dunne et al. 2009; Gomez et al. in preparation) suggests that they could be present
in other SNRs. 
Second, there are large uncertainties in
dust destruction rates. The dust mass of Cas A may explain the lower
limit on the dust masses in high redshift galaxies, when assuming
conservative dust destruction rates \citep{rho08, mor03b}. When
a more moderate destruction rate is taken account, a much higher dust mass per SN is needed
to explain the dust in the early Universe \citep{dwe07,gal08}.  The
grain disruption which occurs primarily in SN-shocks is known to
be too efficient to explain the amount of dust in the ISM \citep{jon94}.
However, new infrared observations of SN 1987A suggest that the
destruction rate of the silicate grains is a factor of 2 smaller than
previous values \citep{dwe08}, and that micro-sized grains can survive
the passage of a SN shock wave \citep{sla04}. Third, an inaccurate
estimation of progenitor masses from ejecta could cause the
discrepancy, since the IMF of massive stars is rather steep. As
discussed in \S4.5, progenitor
mass depends on accurate estimates of ejecta masses over a broad range
of wavelengths.  Fourth, if E0102 contains Fe-bearing dust, the dust
mass would be higher because the density of Fe dust ($\rho$ = 7.95) is
three or more times higher than other grains.  Since Fe dust (due to
the shape of Q$_{abs}$) is featureless like carbon and similar to
that of Al$_2$O3 at low temperatures ($<$ 100 K), it is difficult to
distinguish between Fe, C and Al$_2$O$_3$.  While we assume no Fe dust
because Fe gas lines are not detected, a small contribution of Fe
could significantly increase the dust mass. Fifth, some grain types
may be larger than assumed, which would increase the inferred
mass. Sixth, 
smaller-sized dust  in the ejecta is preferentially destroyed
behind the reverse shock as demonstrated by \citet{noz07}; in the case
of E0102, this has been ongoing for 1000 yrs.
 Deep observations with future telescopes, in particular,
far-infrared/submm observations and theoretical models of 
the time evolution of dust that is formed in the ejecta would be 
needed to settle this issue.

\section{Summary}

Using {\it Spitzer} IRS and IRAC, we have carried out a study of the infrared emission from dust and ejecta in E0102.  Our results may be summarized as follows: 

1.  We have detected infrared emission from E0102 in both continuum and
lines.  These include bright lines of [Ne~ II] (12.8 $\mu$m), [Ne~III]
(15.5 $\mu$m), [O~IV] and/or [Fe~II] (26 $\mu$m) [O~IV] and/or [Fe~II]
(26 $\mu$m), weak lines of [Ne~V] (14.3 $\mu$m), [Ne~V] (24.3 $\mu$m), and
[Ne~III] (36 $\mu$m), and possibly [SII], [SIII] (33.5 $\mu$m), [SiII]
(34.89 $\mu$m).  The ejecta in E0102  are dominated by Ne and O in the
infrared, as in the X-ray and optical.

2. The \neiif\ lines are broadened by 20 -100 \% above the instrumental
resolution, implying a true velocity dispersion of about 2000-4500  km s$^{-1}$.
This is comparable to velocities measured for the optically emitting
ejecta, and indicates that the infrared \neiif\ emission is also from
ejecta.

3. The density of the infrared-emitting Ne ejecta is high.  The
collisionally  excited line diagnostics using \nevf\ and \neiiif\ line
ratios imply that  \neiiif\ lines come from a low ($\sim$600 K) temperature
and [Ne V] lines from high density region.
A shock model with a best velocity parameter of 200 km s$^{-1}$
indicates that the  \nevf\
lines mainly come from the cooling zone, which is hotter dense region, whereas [Ne~II]
and [Ne~III] lines are mainly from the photoinization zone with a  cold
temperature of 400 - 1000 K.

4. We detect a broad dust feature at 18 $\mu$m, that we attribute to
MgSiO$_3$ and Si; the remainder of the dust continuum can be fitted
with either carbon or Al$_2$O$_3$ grains.  The spatial correspondence
of the dust continuum and the ejecta lines indicates that the dust is
freshly formed in the ejecta.

5. The continuum show strong correlation with ejecta of E0102, showing
that dust has formed in the ejecta.
We estimate a dust mass of $7\times 10^{-3}$ to 0.015 M$_\odot$
that is higher than previous estimates by an order of magnitude, but
smaller than that of Cas A. 

6. We estimate the mass of infrared-emitting Ne and O ejecta. 
We estimate infrared emitting Ne ejecta mass of 0.014 M$_\odot$
and a total Ne mass of 1.26 M$_\odot$ derived from the best shock models.
Together with the X-ray masses, 
these imply a progenitor mass of $\sim$30 M$_\odot$.
 
For the future, detailed modeling to obtain accurate ejecta masses for
young remnants such as Cas A and E0102 will be helpful to understand
the production of elements such as Ar deep in the ejecta and their
physical conditions in the remnant.  More broadly, the goal is to
understand the progenitors and the explosions that produced these
remnants. Infrared observations with {\it Spitzer} and {\it ISO} have
allowed us to identify dust features in SNRs and SNe including Cas A
and E0102. However, the presence of cold dust, which can be only
identified in the far-infrared and submm, is still uncertain due to 
limited angular resolution at these wavelengths. In the near-future, we
expect {\it Herschel} observations will significantly advance our
knowledge of cold dust.  Sensitive observations will be very helpful
to find and study the ejecta, so deep infrared mid- and near-infrared
observations with {\it Spitzer} and with future missions such as JWST
will significantly advance our understanding of the questions of
nucleosynthesis and dust formation in supernova explosions.

\acknowledgements
We thank Haley Gomez for insightful, helpful comments on the
manuscript.  J. R. thanks Pierre-Olivier Lagage, Sacha Hony, and Anne
Decourchelle for fruitful discussion  on dust and SN ejecta during
her visit to Saclay. We thank Lee Armus and David Shupe for sharing the
results of the IRS wavelength calibration on the IRS instrumental line
widths, and Steven Finkelstein and William Blair for sharing their
digitalized formats of  HST and the optical images, and Shaun Amy for
an radio image. This work is based on observations made with the
\spitzer\ {\it Space Telescope}, which is operated by the Jet
Propulsion Laboratory, California  Institute of Technology, under NASA
contract 1407. Partial support for this work was provided by NASA through an GO
award issued by JPL/Caltech.

{} 

\newpage
\clearpage

\begin{table} 
\caption[]{Summary of E0102 Properties$^a$}
\label{tproperties}
\vspace{-0.3 cm}
\begin{center}
\begin{tabular}{llllcccc}
\hline \hline
distance & 61.1 kpc (1$'$$\sim$17.7801 pc) & \\
angular size (diameter) & 44$''$ & \\
physical size & 13.04 pc & \\
shock radius & 6.52 pc (22$''$) & \\
reverse shock radius & 4.45 pc (15$''$) & \\
age  &1000 yr, 2050$\pm$600 yr & \\
extinction & A$_{\rm v}$ = 0.08 mag, N$_H$=2$\times 10^{20}$ cm$^{-2}$  \\
\hline
\end{tabular}
\end{center}
\tablenotetext{a}{References are given in the text. } 
\end{table}

\begin{deluxetable}{lllcccccccccccclll}
\tabletypesize{\scriptsize}
\tablewidth{0pt}
\tablecaption{Observed Spectral Line Brightnesses}
\tablehead{
\colhead{Wavelength} & \colhead{Line} & \colhead{FWHM}  & \colhead{Surface Brightness$^a$}  & 
\colhead{Surface Brightness$^b$} & \colhead{A$_{\lambda}$} \\ 
\colhead{ (\mic)} & \colhead{} & \colhead{(\mic)}  & \colhead{(erg s$^{-1}$ cm$^{-2}$ sr$^{-1}$)}  &
\colhead{(erg s$^{-1}$ cm$^{-2}$ sr$^{-1}$)} & \colhead{} 
}
\startdata
   12.8145$\pm$    0.0010&  [Ne~II]       &   0.1529$\pm$ 0.0023  & 5.918E-05$\pm$ 8.774E-07& 5.525E-05&0.0022&\\
   14.3475$\pm$    0.0023&  [Ne~V]        &   0.1797$\pm$    0.0053& 3.974E-06$\pm$ 1.834E-07& 3.633E-06&0.0016&\\
   15.5665$\pm$    0.0014&  [Ne~III]      &   0.2027$\pm$   0.0032& 2.354E-05$\pm$ 1.079E-06& 2.491E-05&0.0019&\\
   24.3410$\pm$    0.0059&  [Ne~V]        &   0.3395$\pm$    0.0142& 2.260E-06$\pm$ 9.676E-08& 2.364E-06&0.0016&\\
   25.9132$\pm$    0.0006&  [O~IV]+[Fe~II]&   0.3430$\pm$    0.0015& 2.077E-05$\pm$ 9.442E-08& 2.116E-05&0.0015&\\
   *33.5160$\pm$    0.0010&  [S~III]       &   0.3489$\pm$0.0040 & 2.394E-06$\pm$3.023E-08 & 2.490E-06&0.0011&\\
   *34.8901$\pm$    0.0010&  [Si~II]       &   0.3459$\pm$0.0027  & 3.429E-06$\pm$ 2.920E-08 & 3.612E-06&0.0010&\\
   36.0783$\pm$    0.0041&  [Ne~III]      &   0.2565$\pm$0.0197 & 4.329E-07$\pm$3.595E-08 & 5.206E-07&0.0009&\\
\enddata
\label{table:tlineflux}
\tablenotetext{a}{Estimated using a Gaussian fitting.}
\tablenotetext{b}{Estimated using integration of line fluxes over wavelength ranges. }
\end{deluxetable}

\begin{table}[!htb]
\scriptsize
\caption[]{Properties of Ne Line-broadening}
\label{tlinebroad}
\vspace{-0.3 cm}
\begin{center}
\begin{tabular}{llllcccc}
\hline \hline
Location &RA. & Dec.  & color & Wavelength (\mic) & FWHM (\mic) & Velocity (km s$^{-1}$) &  Velocity$_{true}^b$ (km s$^{-1}$)\\ 
Center & 01:04:02.05 & -72:01:53.2 & blue &  12.8081$\pm$    0.0057&    0.2235$\pm$    0.0134  & 5226$\pm$304 & 4446$^{+353}_{-361}$  \\ 
 &&&(1)&12.7861$\pm$0.0031 & 0.16733$\pm$0.0070 & 3909$\pm$164 [641$^a$]  \\
 &&&(2)&12.9252$\pm$0.0045 & 0.09391$\pm$0.0068 & 2196$\pm$1640 [2615$^a$] \\
E & 01:04:05.01 & -72:01:55.6 &green &  12.7931$\pm$    0.0012&    0.1465$\pm$    0.0028
 &3433$\pm$64 &2060$^{+104}_{-108}$  \\ 
Scenter & 01:04:02.5 & -72:01:59.6  &red &  12.8104$\pm$    0.0026&    0.1172$\pm$    0.0060
 & 2746$\pm$140\\ 
SE shell  &01:04:04.04 & -72:02:00.5 & Fig. \ref{E0102irs} & 12.8145$\pm$    0.0010 &
0.1529$\pm$ 0.0023 & 3579$\pm$54 & 2295$^{+85}_{-83}$\\
\hline
 \end{tabular}
\end{center}
\vspace{-0.3 cm}
\tablenotetext{a}{Velocity dispersion shift from the line centroid.}
\tablenotetext{b}{Velocity$_{true}$ is estimated velocity dispersion after accounting for 
the instrumental resolution.}
\renewcommand{\baselinestretch}{1}
\end{table}

\begin{deluxetable}{llccccccccccccclll}
\rotate
\tablewidth{0pt}
\tablecaption{Properties of Freshly Formed Dust in E0102}
\tablehead{
\colhead{Model} &
\colhead{$^a$$\Delta \chi^2$ (=$\chi^2$/dof)} &
\colhead{$^b$Dust Compositions (temperature [K], mass [M$_\odot$]) } &
\colhead{ $^f$Mass  (M$_\odot$) }      &
\colhead{  $^f$Total Mass (M$_\odot$)}     
}
\startdata
A &2.41($\sim$398/165)  &{\bf MgSiO$_3$ (116, 3.80E-5), Si (150, 6.20E-5)}, {\it C (55, 5.00E-3)} & $^c$5.12$\times 10^{-3}$ &$^c$ 0.015  \\ 
B &1.98($\sim$389/196)    &{\bf MgSiO$_3$ (117, 3.21E-5), Si (114, 4.4E-4)}, {\it Al$_2$O$_3$ ($\equiv$55,  2.06E-3)} & $^c$2.53$\times 10^{-3}$ & $^c$0.007\\
C &2.18 ($\sim$438/201)    &{\bf MgSiO$_3$ (108, 6.25E-5), Si (142, 9.21E-5)}, {\it Al$_2$O$_3$ ($^d$41, 0.03)} & $^d$0.03& $^d$0.09\\  
D &2.16 (417/193) & {\bf MgSiO$_3$ (124, 2.34E-5), Si (149, 6.3E-5)}, 
& & & \\
 && Al$_2$O$_3$ (60, 3.33eE-4), C (55, 4.350E-3), Mg$_2$SiO$_4$ (100, 1.10E-5) & $^e$4.80$\times 10^{-3}$ & $^e$0.014 \\
\enddata
\label{tbestfit}
\tablenotetext{a}{$\Delta \chi^2$  is the reduced $\chi^2$ and ``dof" is the degrees of freedom.}
\tablenotetext{b}{Compositions in the best fit, where
a few primary compositions are written in bold, and
alternative dust compositions are in italics. A combination of compositions are in a regular font. }
\tablenotetext{c}{The errors are on order of 10-20\%.}
\tablenotetext{d}{The errors could not be constraint.}
\tablenotetext{e}{The errors are on order of 20-40\%.}
\tablenotetext{f}{Mass is the mass within the IRS slit and total mass is the mass estimated over
the entire SNR (see the text for details).}
\end{deluxetable}

{\small
\begin{table}[!htb]
\caption[]{Summary of Ne Column Density and Mass Estimates}
\label{Tneionization}    
\vspace{-0.3 cm}    
\begin{center}    
\begin{tabular}{l|lll|l}    
\hline     
Ne ionization state &
\multicolumn{3}{c|}{from Excitation-rate equation} & from Shock model$^b$\\    
  & Model A$^a$  &
Model B$^a$ & Model C$^a$  & \\
\hline    
 ${\rm [Ne~I]~(cm^{-2})}$    & $2.4\times 10^{15}$     &    
\nodata   & \nodata &  ${\bf 1.5\times10^{15}} $  \\    
${\rm[Ne~II] ~(cm^{-2})}$             & $2.7\times10^{13} $         &    
$4.4\times10^{14}$   & $2.4\times10^{16}$      & ${\bf 7.8\times10^{15}}$ \\    
${\rm[Ne~III] ~(cm^{-2})}$           & $6.8\times10^{12}$         &    
$6.5\times10^{14}$    & $1.5\times10^{15}$      & ${\bf 1.2\times10^{15}}$ \\    
${\rm[Ne~IV] ~(cm^{-2})}$                                  & $9.6\times10^{11}$      &    
\nodata    & \nodata     & ${\bf 8.3\times10^{13}}$ \\    
${\rm[Ne~V] ~(cm^{-2})}$  & $5.9\times10^{11}$          &    
$6.0\times10^{11}$     &   $6.0\times10^{11}$    & ${\bf 1.1\times10^{14}}$ \\    
${\rm [Ne~I-V] ~total ~(cm^{-2})}$  & $7.8\times10^{16}$             &    
$1.1\times10^{15}$    &$2.6\times10^{16}$      & ${\bf 4.0\times10^{16}}$ \\    
\hline
Ne mass$^c$ (M$_{\odot}$) in IRS slit                                    & 0.22             &    
0.003    & 0.07      & ${\bf 0.014}$ \\    
total Ne mass$^c$  (M$_{\odot}$)                               & 0.63               &    
0.01       & 0.21        & {\bf 0.042} \\    
\hline    
\end{tabular}    
\end{center}    
\vspace{-0.3 cm}    
\tablenotetext{a}{Model A: n$_e$=$7\times10^{4}$ cm$^{-3}$, T = 612 K; Model B: n$_e$=$1\times10^{4}$ cm$^{-3}$, T = 350 K;
Model C: n$_e$=$1\times10^{3}$ cm$^{-3}$, T = 215 K.  See the text and Figure \ref{nediag} for details.}
\tablenotetext{b}{The best estimates (in bold)  are from our shock model.}
\tablenotetext{c}{Ne Mass is the mass within the IRS slit and total Ne mass is
the mass estimated over the entire SNR.}
\renewcommand{\baselinestretch}{1}
\end{table}
}

\begin{table}
\caption[]{Comparison of Elements detected$^a$ in E0102 and Cas A}
\label{tcompe0102casa}
\vspace{-0.3 cm}
\begin{center}
\begin{tabular}{l|l|llllllllll|l}
\hline \hline
Wavelength         & SNR &  O   & Ne  &  Mg  & Si    & S     & Fe & Ar & Ca &H & He &references$^a$\\
         \hline
infrared &E0102 &  Y  & Y & \nodata   & U   & U  & pN & N  & \nodata  &N & N & this paper \\
         & Cas A & Y      & Y & \nodata   & \nodata   & \nodata & Y      & Y      & \nodata  & \nodata & \nodata & 1, 2  \\ 
         \hline     
optical &E0102  &  Y & Y& Y & pN  &pN&pN&N & N &N& N &3\\
        & Cas A & Y    &   U  &   U  &  U     & Y   & N   & Y   & N      &Y$^*$   & \nodata & 4, 5 \\
\hline
X-rays  &E0102  &  Y   & Y   & Y  & Y  &U    & U   &  N  & Y   &\nodata &\nodata& 6 \\
  &Cas A  &  Y  & Y   & Y   & Y  &Y   & Y  &  Y & \nodata   &\nodata &\nodata&7\\
\hline
\end{tabular}
\end{center}
\tablenotetext{a}{Y: detected, U: uncertain, pN: probably not detected, N: not detected,
 * indicates that the emission is likely from circumstellar medium instead of from ejecta (see the text for details).}
\tablenotetext{b}{(1) Ennis et al. (2007), (2) Douvine et al. (2000), (3) Blair et al. (200), (4) Fesen et al. (2006),
(5) Fesen et al. (1996),
(6) Flanagan et al. (2004), (7) Hwang, Holt \& Petre (2000)}
\renewcommand{\baselinestretch}{1}
\end{table}

\begin{table}
\caption[]{Comparison between the Ejecta Masses of the SNR E0102 and Nucleosynthesis Yields}
\label{yields}
\begin{center}
\begin{tabular}{l|lll|lll|l}
\hline
  & \multicolumn{3}{c|}{Ne (M$_{\odot}$)}  & \multicolumn{3}{c|}{O (M$_{\odot}$)} & Ar (M$_{\odot}$)  \\
 \hline
 IR ejecta mass & \multicolumn{3}{c|}{{\bf 0.042} (from Shock model)} & \multicolumn{3}{c|}{ 0.78 } & \nodata\\
 X-ray ejecta mass &\multicolumn{3}{c|}{ 0.1 - 2$^a$ ({\bf 0.84}$^b$)} & \multicolumn{3}{c|}{0.3 - 6$^a$} &\nodata\\
total mass of IR and X-ray ejecta& \multicolumn{3}{c|}{0.14 - 2.4 ({\bf 0.882}$^b$)}  & \multicolumn{3}{c|}{1.08 - 6.78 }& \nodata\\
\hline
Nucleosynthesis Models$^c$ & A     & B   & C   & A  & B & C &A\\
13 M$_{\odot}$ model &  0.028 & 0.023 & 0.064 & 0.21 & 0.151 & 0.308 &0.0028 \\
15 M$_{\odot}$ model & 0.039 &0.033 &0.049 & 0.433 &0.368 & 0.520 &0.0063 \\
20 M$_{\odot}$ model &  0.257 &0.191& \nodata&  1.480 &0.800&\nodata   & 0.004 \\
25 M$_{\odot}$ model & {\bf 0.651} &{\bf 0.594} &0.565  & {\bf 3.000} &{\bf 2.997} &1.386 & 0.007 \\
35 M$_{\odot}$ model &    \nodata& \nodata&  {\bf   1.366 }       & \nodata& \nodata&{\bf 3.891} & 0.040 \\
40 M$_{\odot}$ model &  \nodata&{\bf 0.720} &\nodata & \nodata &9.110&\nodata& \nodata \\
\hline
\end{tabular}
\end{center}
\tablenotetext{a}{ From Flanagan et al. (2006).}
\tablenotetext{b}{ Estimates from our shock model.
 Note that our shock model
matches the infrared lines. }
\tablenotetext{c}{The model nucleosynthesis yields are taken from
\cite{thi96} (A), \cite{nom97} (B), and \cite{chi04} (C).
 The best estimate and matching
nucleosynthesis yields are written in bold.}
\renewcommand{\baselinestretch}{1}
\end{table}

\newpage
\clearpage

\newpage
\clearpage

\clearpage

\clearpage

\clearpage

\end{document}